\documentclass[12pt]{iopart}
\usepackage{iopams}
\usepackage{graphicx}
\usepackage[usenames,dvipsnames]{xcolor}
\usepackage{hyperref}
\usepackage[compress,numbers,square,comma]{natbib}
\usepackage{listings}
\usepackage{xspace}
\usepackage{enumitem}
\setlist[enumerate]{topsep=0pt,itemsep=0pt,partopsep=0pt,parsep=0pt}

\hypersetup{
    colorlinks=true,
    linkcolor=cyan,
    filecolor=magenta,      
    urlcolor=blue,
    citecolor=blue,
}

\usepackage{xcolor}
\definecolor{codegreen}{rgb}{0,0.6,0}
\definecolor{codegray}{rgb}{0.5,0.5,0.5}
\definecolor{codepurple}{rgb}{0.58,0,0.82}
\definecolor{backcolour}{rgb}{0.95,0.95,0.92}

\lstdefinestyle{mystyle}{
    backgroundcolor=\color{backcolour},
    commentstyle=\color{codegreen},
    keywordstyle=\color{magenta},
    numberstyle=\tiny\color{codegray},
    stringstyle=\color{codepurple},
    basicstyle=\ttfamily\footnotesize,
    breakatwhitespace=false,
    breaklines=true,                 
    captionpos=b,                    
    keepspaces=true,     
    numbers=left,
    numbersep=5pt,               
    showspaces=false,                
    showstringspaces=false,
    showtabs=false,
    tabsize=2
}

\newcommand{\MLST}{{\it Mach. Learn.: Sci. Technol.} }
\begin{document}

\title[AI Framework for Interpretable Prediction of Molecular and Crystal Properties]
{End-to-end AI Framework for Interpretable Prediction of Molecular and Crystal Properties}

\author{Hyun Park$^{1,3,8}$, Ruijie Zhu$^{2,3}$, E.~A. Huerta$^{3,4,5}$, Santanu Chaudhuri$^{3,6}$, Emad Tajkhorshid$^{1,7,8}$ and Donny Cooper$^{9}$}

\address{$^1$ Theoretical and Computational Biophysics Group, Beckman Institute for Advanced Science and Technology, University of Illinois at Urbana-Champaign, Urbana, Illinois 61801, USA}
\address{$^2$ Department of Materials Science and Engineering, Northwestern University, Evanston, Illinois 60208, USA}
\address{$^3$ Data Science and Learning Division, Argonne National Laboratory, Lemont, Illinois 60439, USA}
\address{$^4$ Department of Computer Science, The University of Chicago, Chicago, Illinois 60637, USA}
\address{$^5$ Department of Physics, University of Illinois at Urbana-Champaign, Urbana, Illinois 61801, USA}
\address{$^6$ Multiscale Materials and Manufacturing Lab, University of Illinois Chicago, Chicago, Illinois 60607, USA}
\address{$^7$ Department of Biochemistry, University of Illinois at Urbana-Champaign, Urbana, Illinois 61801, USA}
\address{$^8$ Center for Biophysics and Quantitative Biology, University of Illinois at Urbana-Champaign, Urbana, Illinois 61801, USA}
\address{$^9$ Computational Science and Engineering, Data Science and AI Department, TotalEnergies EP Research \& Technology USA, LLC, Houston, Texas 77002 USA}

\vspace{10pt}
\begin{indented}
    \item[]\today
\end{indented}

\begin{abstract}
We introduce an end-to-end computational framework 
that allows for hyperparameter optimization 
using the \texttt{DeepHyper} library, 
accelerated model training, and interpretable 
AI inference. The framework is based on state-of-the-art AI models
including \texttt{CGCNN}, 
\texttt{PhysNet}, \texttt{SchNet}, \texttt{MPNN}, 
\texttt{MPNN-transformer}, and \texttt{TorchMD-NET}. 
We employ these AI models along with the benchmark \texttt{QM9}, \texttt{hMOF}, 
and \texttt{MD17} datasets to showcase how the models can predict user-specified material properties within modern 
computing environments. We demonstrate 
transferable applications in the modeling 
of small molecules, inorganic crystals and nanoporous metal organic 
frameworks with a unified, standalone framework. 
We have deployed and tested this framework
in the ThetaGPU supercomputer 
at the Argonne Leadership Computing Facility, 
and in the Delta supercomputer at the National Center for 
Supercomputing Applications to 
provide researchers with modern tools to 
conduct accelerated AI-driven 
discovery in leadership-class computing environments. 
We release these digital assets as 
open source scientific software in GitLab, and 
ready-to-use Jupyter notebooks in Google Colab.
\end{abstract}


\submitto{\MLST}
\maketitle

\section{Introduction}

With the explosion of AI models \cite{schutt2018schnet,unke2019physnet,tholke2022torchmd,klicpera2020directional,xie2018crystal} developed to predict various material properties 
over the recent years, it has become difficult to keep track of 
the available AI models and the datasets that are used for training 
and inference. 
Numerous efforts~\cite{DIG,fung2021benchmarking} have 
been made toward the integration of AI models and 
their associated datasets in one place to streamline 
their use for a wide range of applications and a broad community of 
users~\cite{2022arXiv220700611R,2020arXiv200308394H,gw_nat_ast}. 
AI models and datasets are often 
available through open repositories, in the best scenario, 
so a user can download, deploy and reproduce 
their putative capabilities. Unfortunately, this is a time-consuming 
and laborious process, which can be further complicated when tools and 
libraries used to develop the AI models are 
not available, deprecated, or non-backwards compatible 
in computing environments of new users. 
Furthermore, most of the existing packages are specialized in predicting quantum mechanical properties of small molecules, few of them support crystals.

In order to address these shortcomings, 
here we report the construction 
of a computational framework that 
consolidates libraries, AI models 
and AI interpretability tools to study  
molecules, crystals, and metal-organic frameworks. 
The framework enables hyperparameter tuning 
through the open source library
\texttt{DeepHyper}~\cite{balaprakash2018deephyper}, 
model training, and interpretable inference of 
small-molecule quantum mechanics (QM) properties from 
public datasets such as 
\texttt{QM9}~\cite{ruddigkeit2012enumeration},
and crystal properties from datasets such as 
\texttt{hMOF}~\cite{wilmer2012structure}. 
Key aspects of this computational framework include:

\vspace{4mm}\textbf{Novel features of AI models.} The node and 
edge embedding schemes of two graph neural 
network models, \texttt{PhysNet}~\cite{unke2019physnet} and \texttt{CGCNN}~\cite{xie2018crystal}, 
were modified from 
the original adjacency matrix format to an adjacency list format to reduce redundant information and enable faster 
training. We also adapted small-molecule 
property prediction models to take in crystal structures 
as input such as the crystal version of \texttt{SchNet}~\cite{schutt2018schnet}. 

\vspace{4mm}\textbf{Transferable AI applications.} 
We demonstrate the transferability of the learned force 
fields by training \texttt{TorchMD-NET}~\cite{tholke2022torchmd} model using selected molecular dynamics (MD) trajectory data of a given set of molecules in the
\texttt{MD17} dataset~\cite{chmiela2017machine} to
perform MD simulations of similar molecules. In particular, we show that a model 
trained based on ethanol is transferable to both \textit{n}-propanol and 
iso-propanol, and a model trained based on uracil is transferable 
to pyrimidine and naphthalene. All of the results are 
automatically logged to weights and 
biases (WandB)~\cite{wandb}, a machine learning 
 tracking tool, for simple access.

\vspace{4mm}\textbf{Interpretable AI inference.} 
Intrepretation methods provide a novel pathway to deepen the understanding of the structure-property relationships of materials. Previous work ~\cite{pope2018discovering} has shown great success in applying interpretation methods to identify key functional groups in molecules that contribute to toxicity, including Excitation Backpropagation~\cite{zhang2018top}, CAM~\cite{zhou2016learning} and Grad-CAM~\cite{selvaraju2017grad} and Contrastive gradient~\cite{tieleman2008training}. Moreover, The UMAP method has also been applied to effectively visualize the distribution shift of sampled molecules on a 2D plane with and without transfer learning~\cite{moret2020generative}. To gain a better understanding of model predictions, we provide two interpretation methods to explain the learned features. The first method, Grad-CAM, highlights selected atoms of molecules that are significant for model predictions. 
The second method, uniform manifold approximation and projection~\cite{mcinnes2018umap}, or UMAP, maps the last hidden layer of the model onto a 3D plane. In this way, we can make more sense of the molecular clusters with similar properties.

\vspace{5mm}
This AI suite and scientific software are released 
with this manuscript, and may be found in \texttt{GitLab}~\cite{gitlab_code_md}. To facilitate the use 
of these resources, we have prepared Jupyter notebooks in 
Google Colab~\cite{colab_tutorial_md}, which have been 
tested independently by researchers not involved in this 
work to ensure that these resources are easy to 
follow and use. To ensure that all the resources used 
in this article are self-contained, we have also 
published the datasets used for these studies in 
Zenodo~\cite{hyun_park_2023_7758490}.
We expect that this collection of  
state-of-the-art graph 
neural networks, transformer models, 
and analysis methods for small molecules and crystals will empower AI practitioners 
to seamlessly perform hyperparameter optimization, 
accelerated training, and 
interpretable AI inference in modern computing environments with a unified, standalone computational framework.

\section{Related work}

Graph neural networks have shown great success for modeling 
molecular and crystal structures. For small molecules, a suite of 
models have been proposed, including 
\texttt{DimeNet}~\cite{klicpera2020directional}, 
\texttt{GemNet}~\cite{gasteiger2021gemnet}, 
\texttt{SphereNet}~\cite{liu2022spherical}, 
\texttt{ComENet}~\cite{wang2022comenet}, 
\texttt{SchNet}~\cite{schutt2018schnet} and 
\texttt{PhysNet}~\cite{unke2019physnet}. 
These models take in atomic coordinates and atomic numbers 
as input, 
and represent atoms as nodes and bonds as edges. Typical 
target properties 
for these models are QM properties of molecules such as 
internal energy, 
heat capacity and zero point vibrational energy (\texttt{ZPVE}). For crystal 
structures, periodic boundary conditions need to be considered, 
therefore crystal graph representations are typically used. Example 
graph neural networks that take in crystal structures 
as input include 
\texttt{ALIGNN}~\cite{choudhary2021atomistic}, 
\texttt{CGCNN}~\cite{xie2018crystal}, 
and \texttt{MEGNet}~\cite{chen2019graph}. These models 
first extract crystal 
graphs from the structures, then generate atom and edge embeddings 
for the center atoms and their neighbors. The bond and edge information 
is then updated via message passing. The target properties for these models are typically QM properties of crystals, e.g., formation energy and band gap.

The growing number of the graph neural networks available for this purpose pushes the need for 
an end-to-end AI framework. Previous efforts toward 
such a goal typically missed one or more important aspects. For example, 
\texttt{MatDeepLearn}~\cite{fung2021benchmarking} integrates 
a suite of 
graph neural networks, including \texttt{CGCNN}, \texttt{MEGNet}, 
\texttt{MPNN}~\cite{gilmer2020message}, \texttt{GCN}~\cite{schlichtkrull2018modeling}
and \texttt{SchNet}. Although it can be used for 
hyperparameter tuning, model 
training, and inference, it lacks the explainability feature, which limits the amount of 
chemical insights that could be extracted from the results. Another 
example is Dive in Graphs (DIG)~\cite{DIG}, which enables model 
training and explanation. However, it does not allow for 
hyperparameter tuning, therefore only models with preset 
hyperparameters can be used. A complete package 
offering all of the aforementioned functionalities is therefore needed.

Our AI framework also offers the functionality to 
perform MD simulations for small molecules, enabled by 
\texttt{TorchMD-NET}, an SE3-equivariant 
transformer interatomic potential model that establishes 
a relationship 
between atomic configurations and potential energies and forces. 
The MD trajectories of selected molecules taken from the \texttt{MD17} dataset were used for training the \texttt{TorchMD-NET} models.

\section{Methods}
\label{sec:met}

Here we describe the key building blocks of our general-purpose AI framework: 

\begin{enumerate}[nosep]
    \item It provides built-in datasets 
    and neural networks that we modified to take in 
    adjacency list format
    node and edge embeddings, a more memory efficient 
    embedding scheme than adjacency matrix format
    \item It enables distributed hyperparameter tuning of 
    neural networks via the scalable and computationally efficient 
    library \texttt{DeepHyper}
    \item Model training and interpretable inference are performed 
    by specifying a few command line arguments
    \item Results are auto-logged to \texttt{WandB}, a machine 
    learning tool for easy tracking and visualization
    \item Molecular dynamics simulations can be performed for small 
    molecules using \texttt{TorchMD-NET} if trained 
    with MD trajectories from the \texttt{MD17} dataset, enabled 
    by the atomic simulation environment (ASE) library~\cite{larsen2017atomic}.
\end{enumerate}

This framework has been deployed and tested in leadership 
computing platforms to reduce the overhead for researchers 
that require access to hyperparameter tuning, model training and explainable inference tools in a single, unified framework. 
Below we describe each of these components in further detail.

\paragraph{Hyperparameter tuning.} This feature was done using the 
\texttt{DeepHyper}~\cite{balaprakash2018deephyper} library. 
In this method, hyperparameters of interest are given prior 
distributions and their posterior distributions are adjusted 
based on the Centralized Bayesian Optimization (CBO) algorithm with a 
given acquisition function and a surrogate model. The graph neural 
networks in this framework are coupled with \texttt{DeepHyper} to 
enable faster hyperparameter tuning.

\paragraph{Datasets}
\texttt{QM9} and \texttt{MD17} datasets were used as input 
to graph neural networks. 
The \texttt{QM9} dataset consists of molecular structures 
and QM properties of 
133,885 molecules with up to nine atoms of type H, C, O, N and F. 
For demonstration purposes, the selected QM properties in this work include 
the highest occupied molecular orbital (\texttt{HOMO}), and zero 
point vibrational energy (\texttt{ZPVE}). The \texttt{MD17} 
dataset consists 
of ab-initio MD trajectories of 10 molecules at 
different levels of theory. Both datasets are available in the 
\texttt{PyTorch Geometric} library.

\paragraph{Node and edge embedding schemes.} Instead of using 
the original adjacency matrix format for node and edge embeddings, 
we modified it to adjacency list format. 
The term embedding, for 
both molecular and crystal graphs, refers to the information 
attached to a node (an atom) or an edge (a bond). Both node and 
edge embeddings can be scalars, vectors or higher order tensors. 
Node embeddings encode information such as mass, charge and 
orbital hybridization, whereas edge embeddings encode information 
such as interatomic distance and bond order. Depending 
on the model architecture, some embeddings are physics- or 
chemistry-based while others are learned. For physics- or 
chemistry-based embeddings, fixed information such as hybridization, mass, atomic 
radius, and whether the fragment is a part of an aromatic ring is encoded. On the other hand, learned embeddings refer to embeddings that are iteratively optimized by a neural network model via stochastic gradient descent.

In adjacency matrix format edge embeddings, the adjacency matrix 
is encoded into a fixed-size matrix, whose size is determined by 
the largest molecule in the dataset. For other molecules, their vectors 
are padded to be the same dimension as the largest one. Each element 
in the adjacency matrix indicates whether the two corresponding nodes 
are connected, as determined via some distance-based criteria. 
Since padding is applied to smaller molecules in the matrix, users 
need to know a priori the largest molecule size, then perform masking 
to obtain the padding values, which can be burdensome for GPU memories. 
By using the adjacency list format, however, only the information for 
connected atoms is preserved, thereby avoiding the need for padding and taking less memory to load.
In this case, faster loss convergence and higher prediction accuracy are expected. 
The adjacency list format has been implemented in a number of \texttt{Python} 
libraries such as \texttt{Deep Graph Library (DGL)}~\cite{wang2019deep}
and \texttt{PyTorch Geometric}~\cite{Fey/Lenssen/2019}. 
In this work, we use \texttt{CGCNN} model as an example to 
demonstrate a boost in model training performance 
when an adjacency list format is used in place of an adjacency matrix format. 

Our AI framework allows users to perform hyperparameter tuning, 
model training and interpretable inference for pre-trained models 
or train new models with a few arguments passed. The main improvements 
over previously proposed general-purpose machine learning model libraries 
is the explainability feature, which consists of two parts. First, by 
extracting high dimensional hidden layer information from the 
learned models and projecting it onto low dimensions via the UMAP method, 
we can effectively visualize the clustering of molecules, 
with similar practice 
as in~\cite{leow19GraphTSNE,gelman2021neural,mnih2015human}. 
Second, Saliency Map~\cite{simonyan2013deep}, CAM and Grad-CAM methods are used to highlight important atoms in molecular graphs, as described in~\cite{pope2019explainability}.

\section{Results}

Below we present a comprehensive analysis of 
our results, from hyperparameter optimization to 
interpretable AI inference.

\subsection{Hyperparameter Optimization}
\label{sec:hptmb}

The \texttt{DeepHyper} library is used for hyperparameter optimization of 
graph neural networks. \texttt{DeepHyper} is easy to use and can 
be readily deployed on GPU-based high-performance computing platforms. 
CPUs can be used if GPUs are not available. However, if the user 
has access to multiple GPUs, then the GPU option will be automatically 
chosen, with each core performing hyperparameter search using the CBO algorithm, given an 
acquisition function such as the upper confidence bound, 
and a surrogate model, e.g., random forest.

The list of hyperparameters considered in 
this work along with their ranges are summarized in 
Table~\ref{tab:params}. 
Hyperparameter optimization results for \texttt{PhysNet} with 
\texttt{ZPVE} as target property are shown 
in Tables~\ref{tab:params_a1} and~\ref{tab:params_a2}. 
It is worth mentioning that 
since \texttt{DeepHyper} tries to 
maximize the objective of search, the opposite number of 
validation error was used as the objective, therefore a 
larger absolute value of objective corresponds to a
better combination of hyperparameters. 
The hyperparameter tuning results for \texttt{PhysNet} with \texttt{HOMO} as the target property are shown in Tables~\ref{tab:params_ap1} and~\ref{tab:params_ap2}. For hyperparameters with integer or floating number values, the ranges represent the lower and upper bounds. For hyperparameters \texttt{amp} and \texttt{optimizer}, the ranges represent all available options.

\begin{table}[htbp]
\centering
\small
\caption{List of hyperparameters and their ranges.}
\begin{tabular}{ccc}
\hline
      hyperparameter & log scale & range \\ \hline
      \texttt{agb} & true & [1,20] \\ \hline
      \texttt{amp} & false & [true,false] \\ \hline
      \texttt{batch size} & / & [128,512] \\ \hline
      \texttt{epochs} & true & [10,100] \\ \hline
      \texttt{gradient clip} & true & [1e-05,2] \\ \hline
      \texttt{learning rate} & true & [1e-3,1] \\ \hline
      \texttt{optimizer} & / & [SGD,TorchAdam,Adam,LAMB] \\ \hline
      \texttt{weight decay} & true & [2e-6,0.02] \\ \hline
    \end{tabular}
    \label{tab:params}
\end{table}

Among the hyperparameters, \texttt{agb}, or accumulated 
grad batches, helps overcome memory constraints; 
\texttt{amp}, or automatic mixed precision, speeds up neural network training; and \texttt{gradient clip}, a machine learning technique where the gradients of neural network parameters are rescaled to between 0 and 1,
is known to stabilize neural network training 
by avoiding sudden changes in parameter values (also known 
as the exploding gradient problem)~\cite{GCLIP}.

\begin{table}[htbp]
    \centering
    \small
    \caption{Top 10 \texttt{DeepHyper} hyperparameter combinations 
    for \texttt{PhysNet} with \texttt{ZPVE} as target property.}
    \begin{tabular}{cccc}\\
    \hline
\texttt{agb} & \texttt{amp} & \texttt{batch\_size} & \texttt{gradient\_clip}  \\ \hline
4  & TRUE  & 190 & 0.00245 \\
3  & TRUE  & 190 & 0.00022 \\
1  & TRUE  & 397 & 0.00032 \\
4  & TRUE  & 196 & 0.00122 \\
3  & TRUE  & 174 & 1.60E-05 \\
4  & TRUE  & 154 & 3.25E-05 \\
4  & TRUE  & 300 & 0.732 \\
4  & TRUE  & 228 & 0.00389 \\
2  & FALSE & 359 & 0.72537 \\
11 & TRUE  & 168 & 0.00243 \\ \hline
    \label{tab:params_a1}
    \end{tabular}
\end{table}

\begin{table}[htbp]
    \centering
    \small
    \caption{As Table~\ref{tab:params_a1} for the rest of parameters 
    optimized through \texttt{DeepHyper}.}
    \begin{tabular}{cccc}\\
    \hline
\texttt{learning\_rate} & \texttt{optimizer}   & \texttt{weight\_decay} & \texttt{objective}  \\ \hline
0.00296 & torch\_adam  & 1.03E-05 & -0.9226 \\
0.75169 & torch\_adam  & 5.14E-06 & -6.7526 \\
0.00015 & lamb         & 2.69E-06 & -6.7925 \\
0.32274 & lamb         & 1.45E-05 & -7.1168 \\
0.17673 & lamb         & 7.80E-06 & -12.31  \\
0.00144 & torch\_adam  & 1.13E-05 & -15.394 \\
0.02986 & lamb         & 3.57E-06 & -26.13  \\
0.00966 & torch\_adam  & 7.16E-06 & -27.627 \\
0.02491 & sgd          & 1.04E-05 & -29.335 \\
0.00124 & torch\_adamw & 0.00011  & -30.203 \\ \hline
    \label{tab:params_a2}
    \end{tabular}
\end{table}

\noindent The optimal hyperparameter combinations found by \texttt{DeepHyper} 
are listed in the top rows of Tables~\ref{tab:params_a1} 
and~\ref{tab:params_a2}, with the optimal objective being -0.9226. 
We notice that this set of hyperparameters include f32 
precision (amp="false"), a standard learning rate 
(0.00296), and a low gradient norm 
clipping value (0.00245). These result in small gradient accumulation, 
which may help mitigate sudden gradient updates.

We have tested multiple sets of hyperparameters with varying ranges 
and prior distributions. Our hyperparmeter tuning configuration 
input file is prepared in \texttt{YAML} format. Discrete hyperparmeter
values such as the number of epochs and the batch size are
sampled from uniform distributions whereas continuous hyperparameters such as the learning rate and the gradient clip 
are sampled from normal distributions with/without log scale. 
The ranges of hyperparameters along with the prior distributions 
for sampling are both user-customizable.

Once the hyperparmeter configuration and the prior 
distributions are in place, \texttt{DeepHyper} can 
use multiple 
GPUs to perform hyperparameter tuning, taking 
full advantage of 
GPU parallelization. Next, all the optimization 
results will be saved 
and automatically logged to Weights and Biases. 
If the tuning step is interrupted, 
it can be resumed from the last saved 
checkpoint by specifying the \texttt{----resume} tag. 

\begin{figure}[ht]
   \centering
   \includegraphics[scale=0.7]{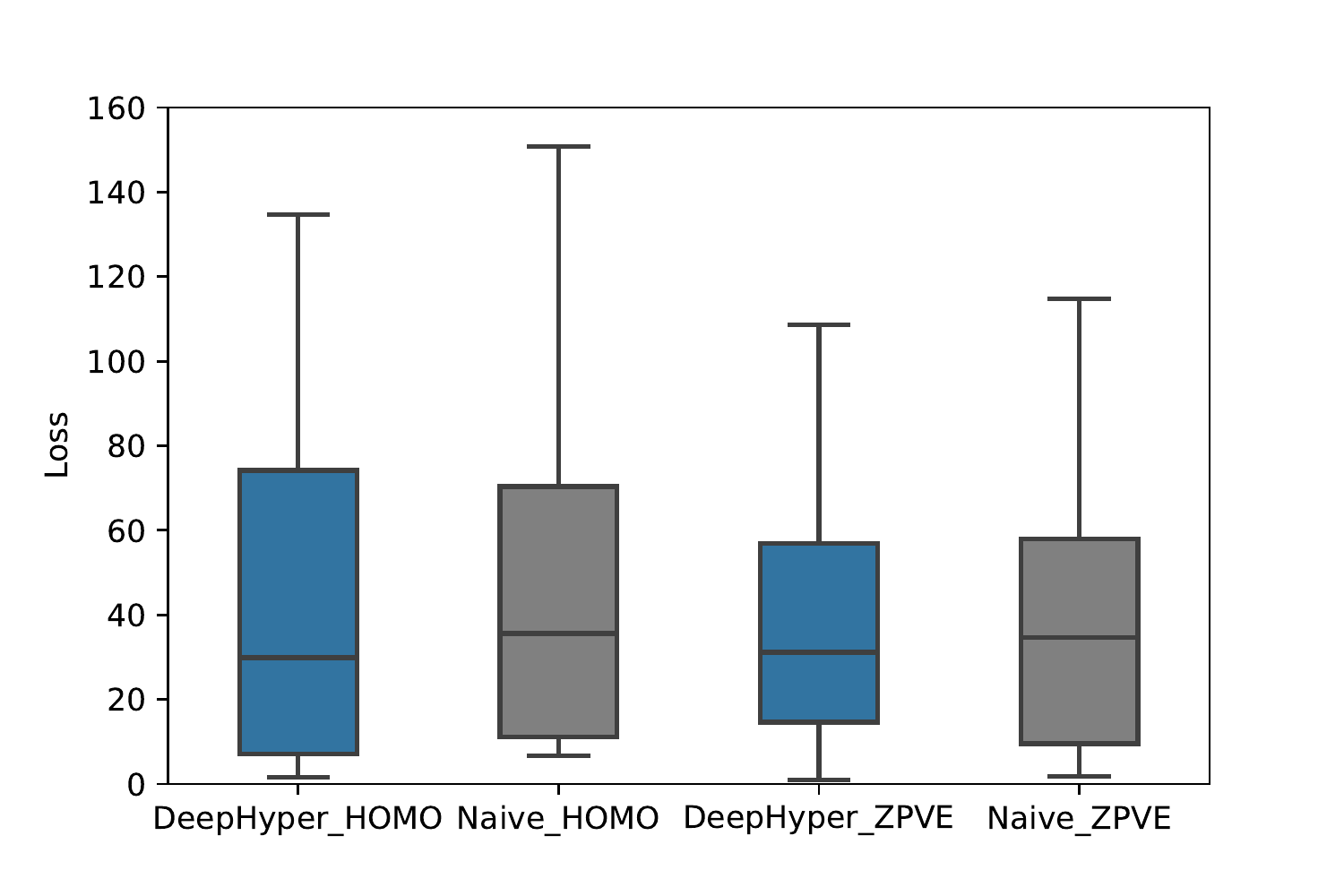}
       \caption{Comparison of the loss distributions of \texttt{PhysNet} with hyperparmaters found by \texttt{DeepHyper} (blue) and a naive random selection algorithm (gray). Two outliers for the \texttt{DeepHyper\_HOMO} box were neglected to retain details.}
   \label{fig:deephyper_vs_naive.pdf}
\end{figure}

For the \texttt{PhysNet} model with \texttt{HOMO} 
and \texttt{ZPVE} as target properties, 
we compared hyperparameter tuning performance 
of \texttt{DeepHyper} with a naive algorithm that 
performs random selection of hyperparameters. 
Since \texttt{DeepHyper} utilizes the CBO algorithm to 
optimize hyperparameters, the target property values 
are used for decision making. For the naive 
algorithm, however, hyperparameters were randomly 
selected from the hyperparameter grid 
in Table~\ref{tab:params}. A total of 20 models 
were trained for 30 epochs with hyperparameters given 
by the two methods. The distributions of the losses (mean squared error)
are compared in Figure ~\ref{fig:deephyper_vs_naive.pdf}, 
and the metrics are summarized in 
Table~\ref{tab:deephyper_vs_naive}.

\begin{table}[ht!]
\centering
\small
\caption{Performance of 20 models with hyperparameters found by \texttt{DeepHyper} and a naive random selection algorithm with \texttt{HOMO} and \texttt{ZPVE} as target properties. For both properties, the minimum loss and the standard deviation of loss are reported.}
\begin{tabular}{ccccc} \hline
                & \multicolumn{2}{c}{HOMO} & \multicolumn{2}{c}{ZPVE} \\ \hline
                & min\_loss   & std\_loss  & min\_loss   & std\_loss  \\ \hline
DeepHyper       & 1.604       & 74.871     & 0.923       & 31.771     \\
Naive Algorithm & 6.632       & 52.43      & 1.751       & 65.117     \\\hline
\label{tab:deephyper_vs_naive}
\end{tabular}
\end{table}

\vspace{3mm}

\noindent \textit{Key findings:} For the prediction 
of \texttt{HOMO} and \texttt{ZPVE}, \texttt{DeepHyper} 
yields better hyperparameter combinations, which accelerate 
convergence and provide optimal performance. 

\subsection{AI model training}

\begin{figure}[ht]
    \centering
    \includegraphics[scale=0.52]{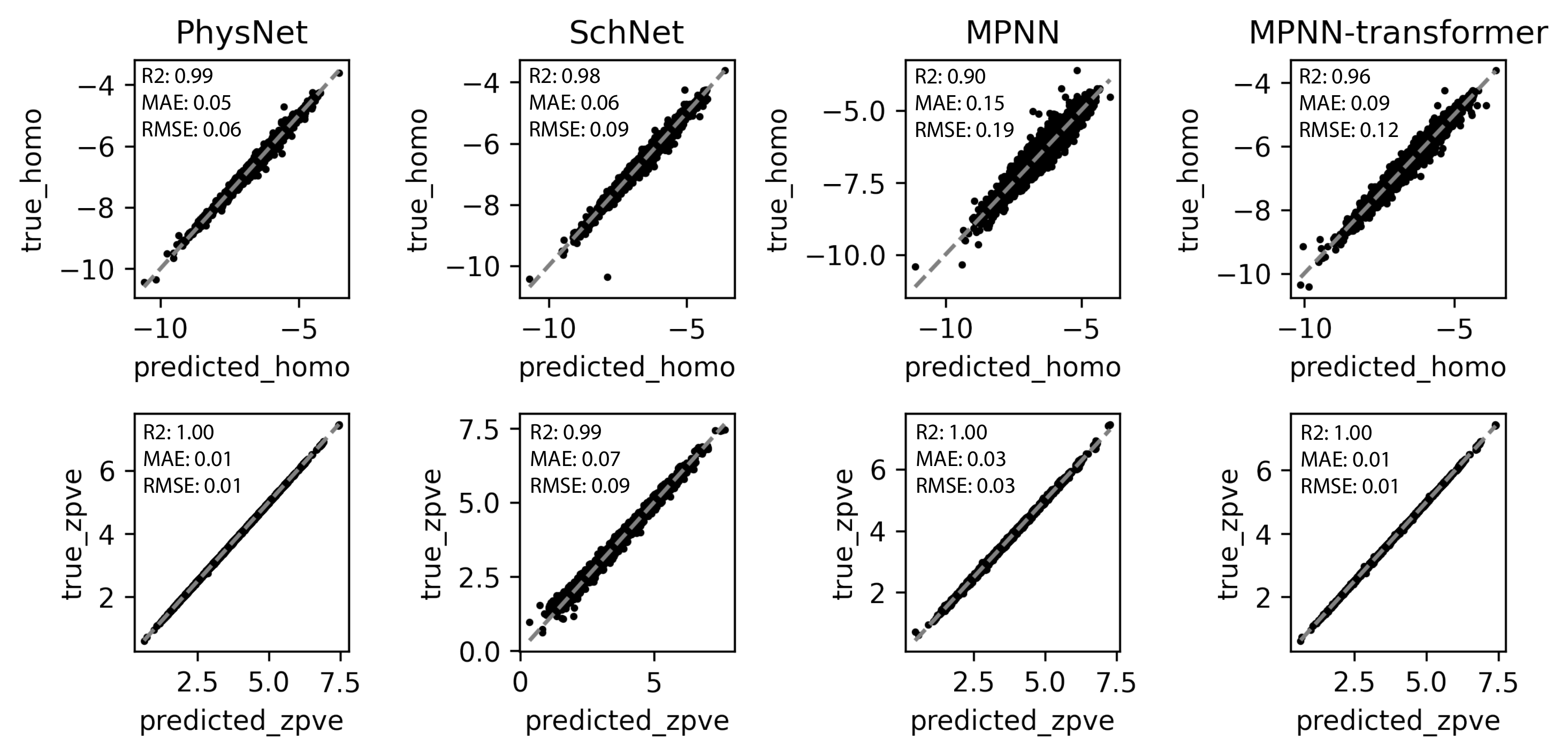}
    \caption{From left to right, model inference performance of 
    \texttt{PhysNet}, \texttt{SchNet}, \texttt{MPNN} and 
    \texttt{MPNN-transformer} with \texttt{HOMO} (top row) or  
    \texttt{ZPVE} (bottom row) as the target property.}
    \label{fig:model_pred}
\end{figure}

We trained \texttt{PhysNet}, \texttt{SchNet}, \texttt{MPNN} and 
\texttt{MPNN-transformer} (with attention mechanism)
with \texttt{HOMO} and \texttt{ZPVE} as target properties from the
\texttt{QM9} dataset. 
The models were trained for 1,500 epochs to ensure convergence of validation loss. 
A new model is saved when the validation loss drops.
The model training results are summarized in 
Figure ~\ref{fig:model_pred}. We found that \texttt{ZPVE} is an 
easier property to learn compared to \texttt{HOMO} for all four
 models, as indicated by significantly lower losses. 
 Moreover, the addition of attention layer in the
\texttt{MPNN} model (\texttt{MPNN-transformer}) further lowers the mean absolute error
(0.09\,eV for \texttt{HOMO} and 0.01 eV for \texttt{ZPVE}) compared to the original \texttt{MPNN} model 
(0.15\,eV for \texttt{HOMO} and 0.03 eV for \texttt{ZPVE}).

\begin{figure}[ht!]
    \centering
    \centerline{
    \includegraphics[scale=0.5]{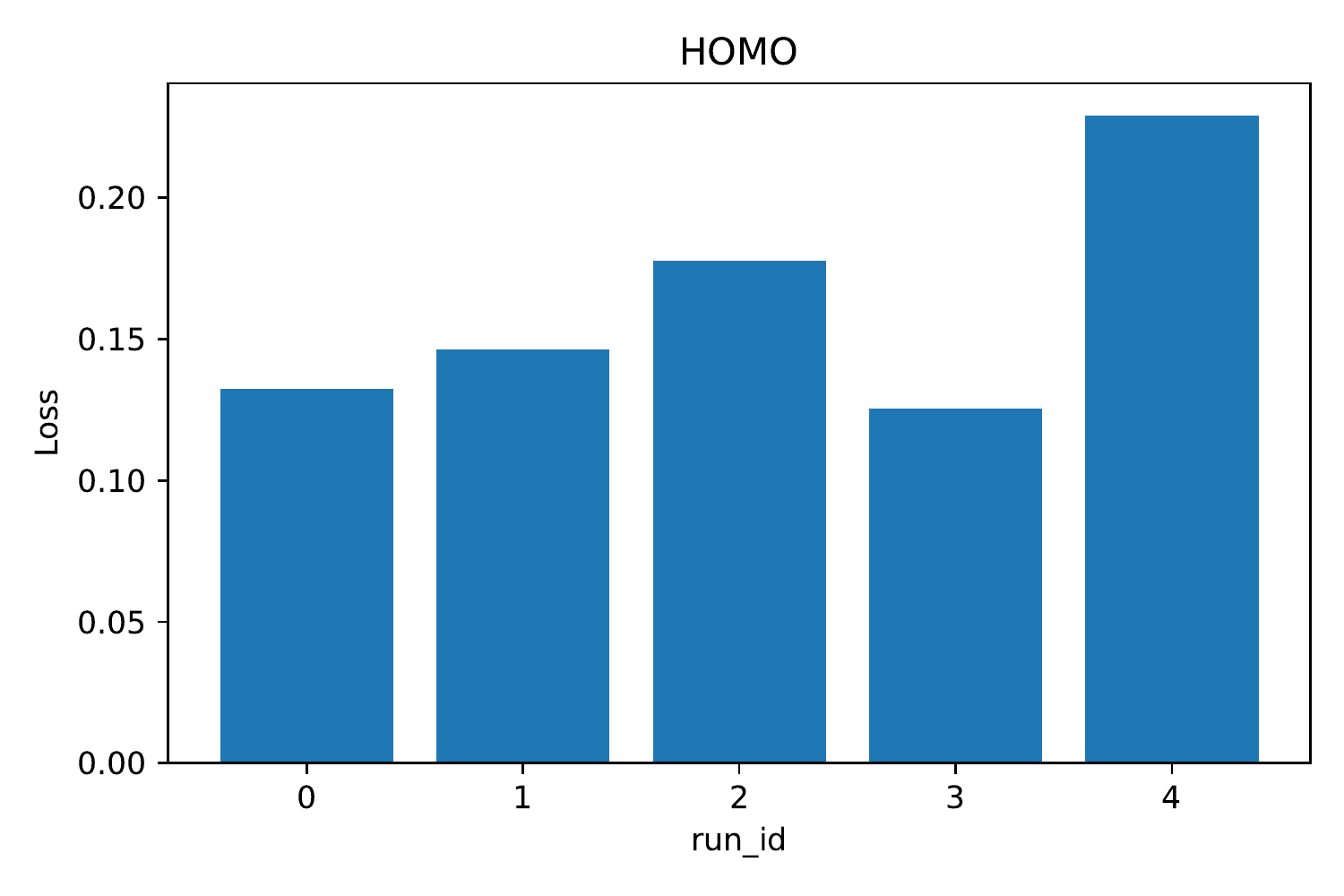}
    \includegraphics[scale=0.5]{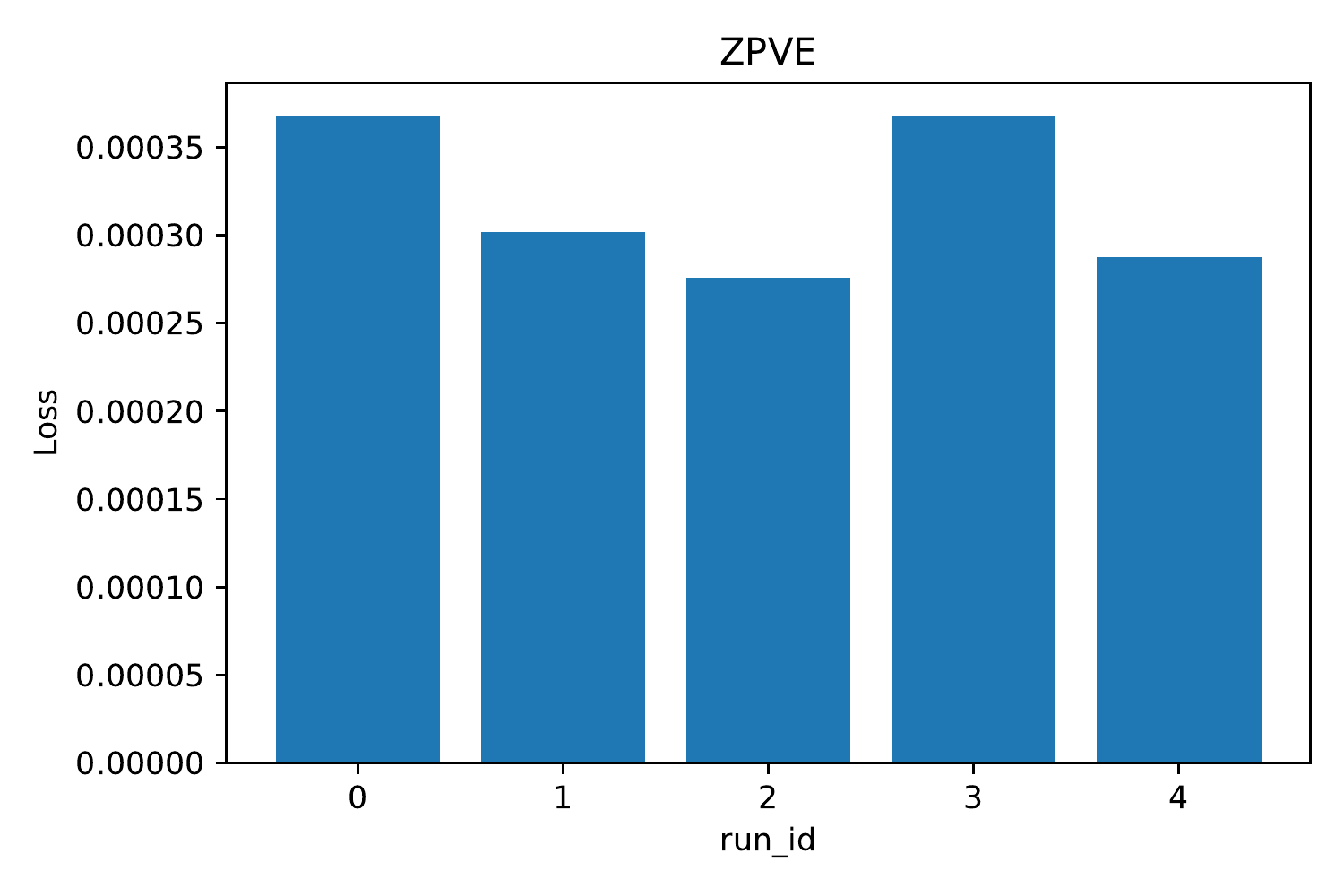}
    }
        \caption{Model training performance of \texttt{PhysNet} with \texttt{HOMO} (left) and \texttt{ZPVE} (right) as target properties, initialized with 5 random seeds.}
    \label{fig:uncertainty_zpve.pdf}
\end{figure}

Model uncertainty quantification was performed for \texttt{PhysNet} model with \texttt{HOMO} and \texttt{ZPVE} as target properties. Five \texttt{PhysNet} models with randomly initialized weights were generated using the random seeds method. The optimal hyperparameter combinations found by \texttt{DeepHyper} in Section \ref{sec:hptmb} were used. The models were trained for 100 epochs to achieve convergence of loss function. Figure \ref{fig:uncertainty_zpve.pdf} shows that \texttt{PhysNet} makes consistent predictions regardless of the random initial weights. The standard deviations of losses for the five models with \texttt{HOMO} and \texttt{ZPVE} as target properties are 0.0379 eV and 3.9646e-05 eV, respectively, and the mean absolute errors are comparable with those reported in the literature~\cite{HOMO-liter}.

\vspace{3mm}

\noindent \textit{Key findings:} Our suite of AI models provide 
state-of-the-art results. Novel features that we added to the 
models, such as attention in \texttt{MPNN-transformer}, further 
improve model performance. We have also demonstrated that 
hyperparameter optimization leads to stable, statistically robust 
AI predictions. 

\subsection{Model improvement via modified node and edge embedding schemes}

\begin{figure}[htbp]
    \centering
    \centerline{
    \includegraphics[scale=0.54]{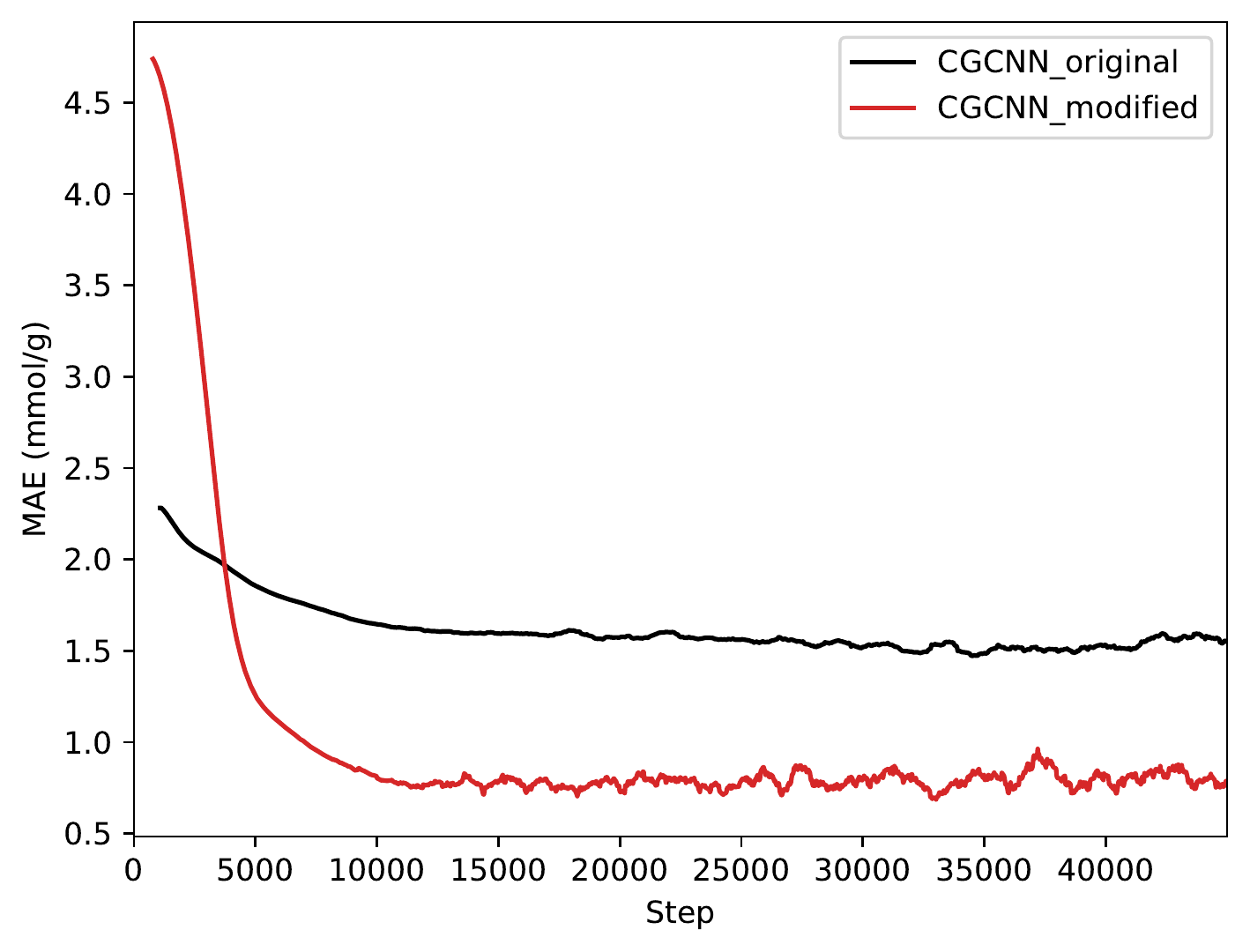}
    \includegraphics[scale=0.51]{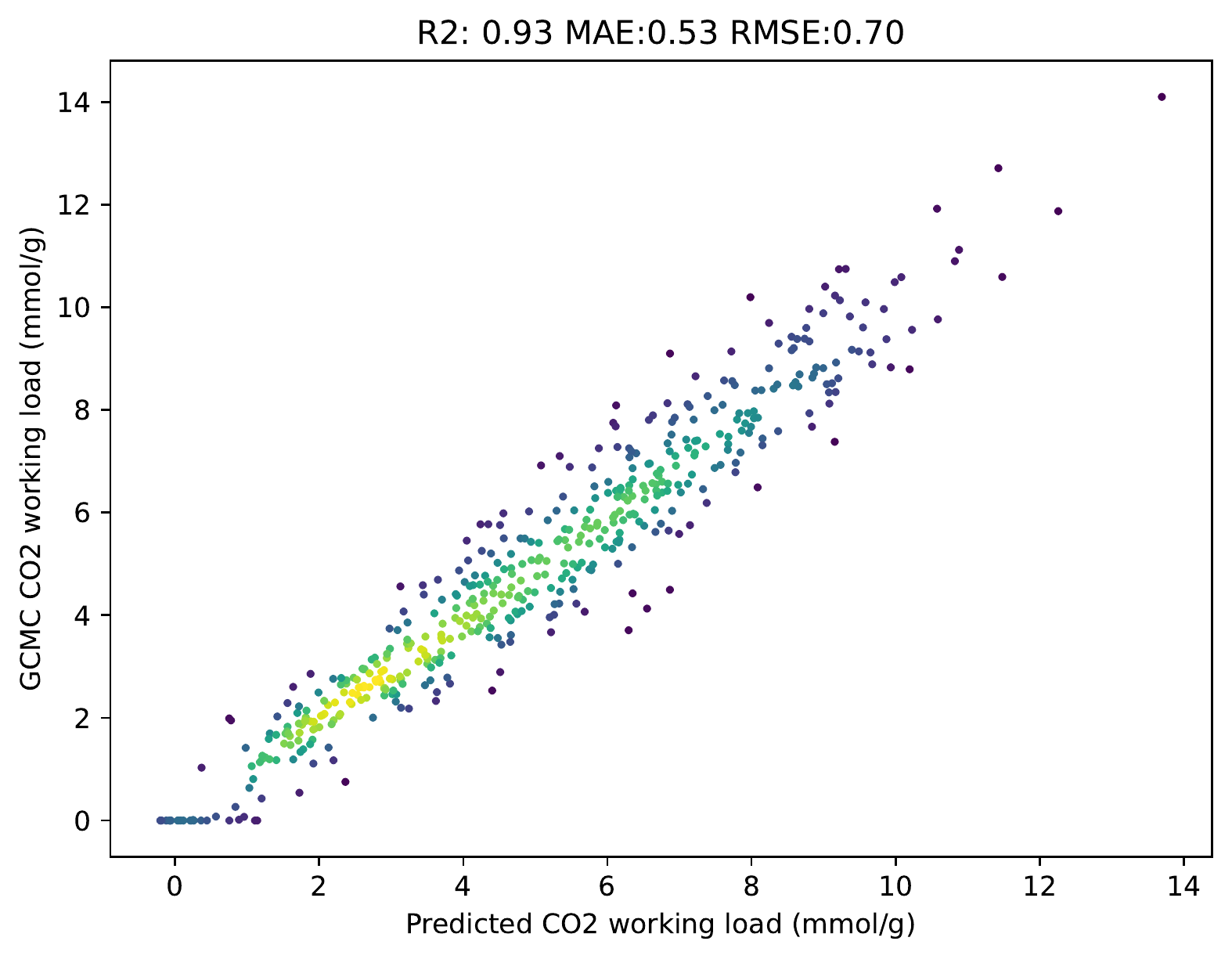}
    }
    \caption{Left panel, comparison of the original \texttt{CGCNN} model (black) with adjacency matrix format node and edge embedding schemes and the modified \texttt{CGCNN} model (red) with adjacency list format node and edge embedding schemes. Right panel, inference performance of the modified \texttt{CGCNN} model on a test set of 500 MOFs of the 
    \texttt{hMOF} dataset.}
    \label{fig:graph_vs_ori}
\end{figure}

We modified the node and edge embedding schemes of 
\texttt{CGCNN} model from
the original adjacency matrix format to an adjacency list 
format. 
There are two main advantages in using the adjacency list 
format. First, compared to the adjacency matrix format, 
it takes up less memory for loading, which speeds up
model training. Second, the redundant information 
(zero paddings) in the representation is removed, resulting in higher training accuracy and stability. As an example, \texttt{CGCNN} 
models with the two embedding schemes were trained on a subset of the \texttt{hMOF} database~\cite{hMOF}, which contains 
5,000 randomly selected MOF structures along with their $\textrm{CO}_2$ working capacities at 
2.5 bar. The 5,000 MOFs are splitted into 80\% training set, 10\% validation set and 10\% test set. The mean absolute error on the test set as a function of training steps for both models are shown in left panel of Figure ~\ref{fig:graph_vs_ori}. 
To smooth out local fluctuations, thirty point moving averaging was performed on both curves. We notice that the modified 
\texttt{CGCNN} model achieved faster convergence speed, higher training stability, and a lower mean absolute error compared to the original model. 

From the right panel of Figure ~\ref{fig:graph_vs_ori},
we show that the modified \texttt{CGCNN} model predicts $\textrm{CO}_2$ working capacity with an R$^2$ score of 0.93 and a mean absolute error of 0.53 mmol/g. To better understand the predictive performance of the modified \texttt{CGCNN} model, we benchmarked it against two recently proposed machine learning models for predicting 
$\textrm{CO}_2$ working capacity of MOFs, namely ALIGNN~\cite{MOF_CO2_1} and random forest regressor~\cite{MOF_CO2_2}. \\
When trained on the entire hMOF dataset, ALIGNN predicts $\textrm{CO}_2$ working capacity at 2.5 bar with a mean absolute error of 0.48 mmol/g~\cite{MOF_CO2_1}. ALIGNN uses both normal graph and line-graph embedding schemes for training and inference of working capacity prediction. For a normal graph, it uses physical and chemical features for node embedding, and distances between atoms as edge embedding; for a line graph, distances correspond to nodes whereas bond angles correspond to edges. This scheme, however, can cause occasional CUDA memory issues and training batch size may have to be reduced (64 in \cite{MOF_CO2_1}; 32 in our independent experiment), hence slower training. On the other hand, our modified \texttt{CGCNN} model only takes in crystal structures as input (i.e., atomic species and Cartesian coordinates) with larger batch size (256 in our model).\\
The random forest regression model takes in topological, structural, and word embedding features as input. When trained on the entire hMOF dataset, it achieves an R$^2$ and RMSE score of roughly 0.95 and 0.65, respectively, for the prediction of CO$_2$ working capacity at 2.5 bar.\\
In other papers, extensive physical and chemical featurization schemes were used \cite{burner2020high}\cite{moosavi2022data} to feature the MOF structures, whereas our modified CGCNN model captures MOF information solely from atom species and coordinates. \\
It is worth noting that we trained the modified CGCNN model on 5,000 randomly selected MOFs instead of all of the structures, therefore lower prediction error is expected if the model is trained on the entire hMOF dataset. Overall, the modified \texttt{CGCNN} model achieves competitive predictive performance compared to state-of-the-art machine learning models.

\begin{figure}[htbp]
    \centering
    \includegraphics[scale=0.45]{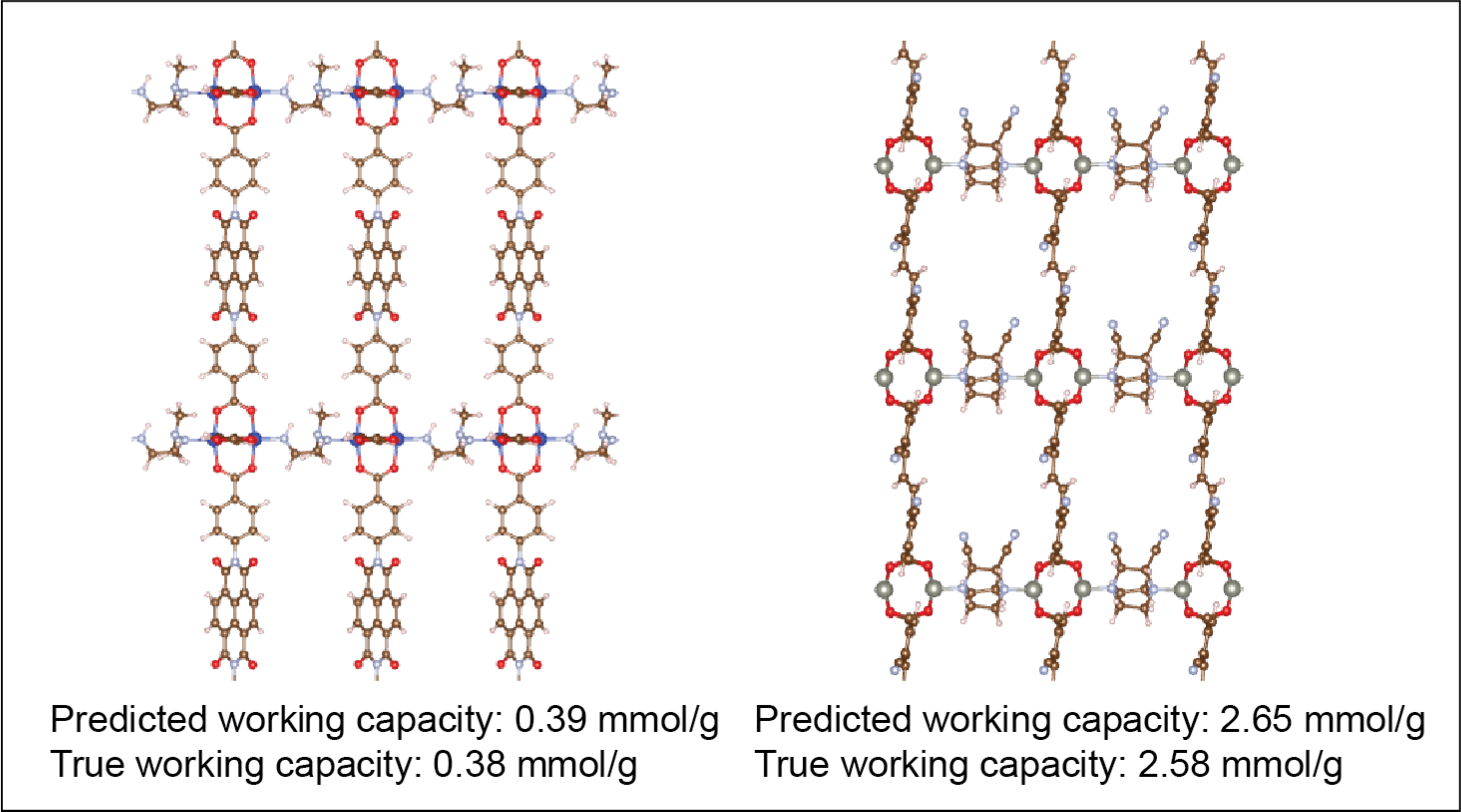}
    \caption{Sample MOF structures in the \texttt{hMOF} 
    database along with the AI predicted and ground 
    truth $\textrm{CO}_2$ working capacities at 2.5 bar.}
    \label{fig:MOF_example} 
\end{figure}

\vspace{3mm}

\noindent \textit{Key findings:} Adopting adjacency list format node and edge embedding scheme improves the predictive capabilities 
of our modified \texttt{CGCNN} model. When making inference on a test set of 500 MOFs from the \texttt{hMOF} dataset, we have found 
that our modified \texttt{CGCNN} model provides 
state-of-the-art predictions for $\textrm{CO}_2$ 
working capacities at 2.5 bar.

\subsection{Transferable AI Applications}
Molecular dynamics simulations of two sets of small molecules 
were performed to demonstrate the transferability of 
\texttt{TorchMD-NET}: 
from ethanol to \textit{n}-propanol/\textit{iso}-propanol, and from uracil to pyrimidine/naphthalene. 

\begin{figure}[htbp]
    \centering
    \includegraphics[scale=0.3]{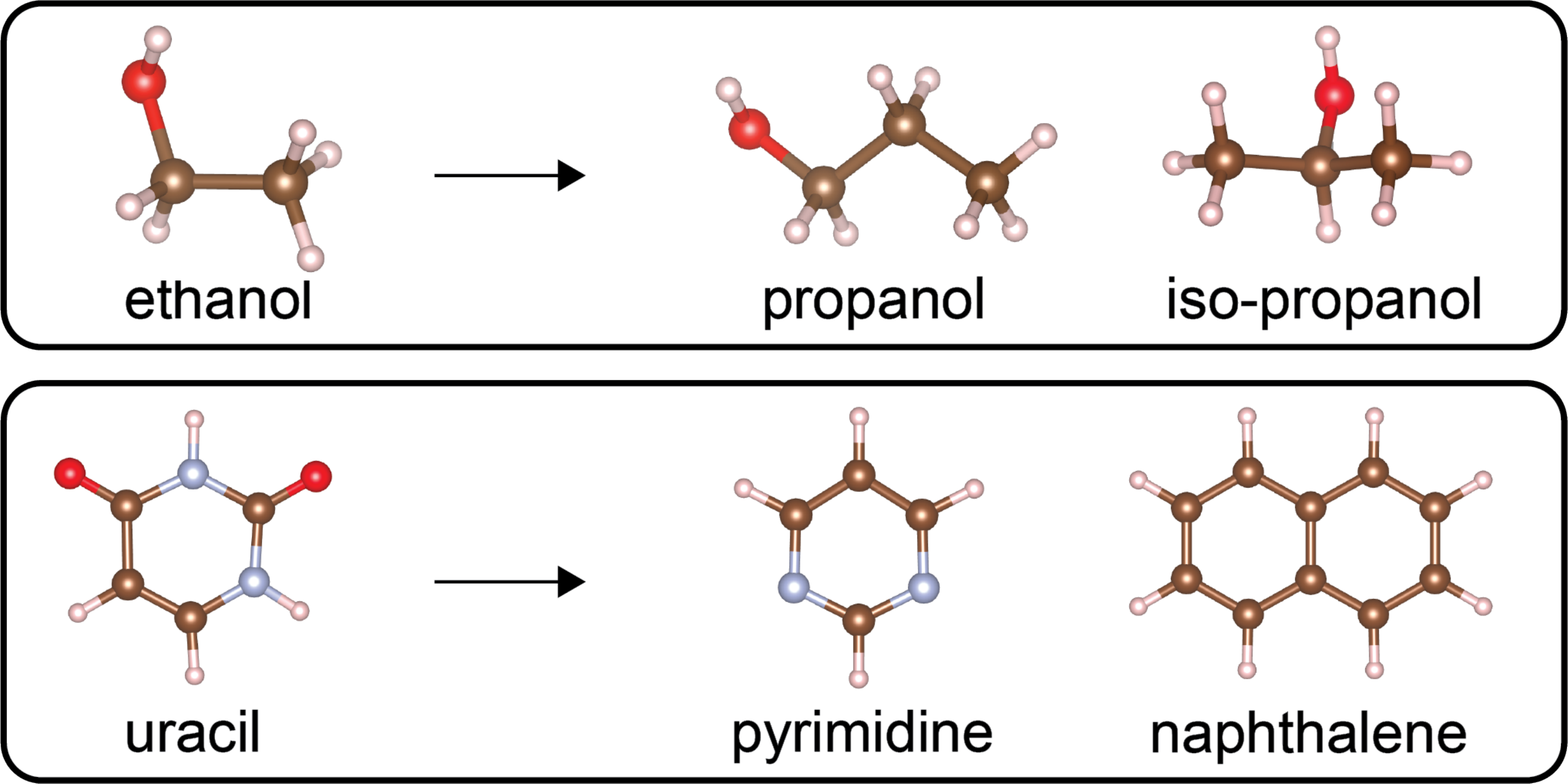}
    \caption{Example molecules used to demonstrate the transferability of \texttt{TorchMD-NET}. MD trajectories of molecules on the left are used to train the \texttt{TorchMD-NET} models, which are then used to perform MD simulations for the molecules on the right.}
    \label{fig:molecular_graph}
\end{figure}

\begin{figure*}[ht]
\centering
\includegraphics[width=1\linewidth]{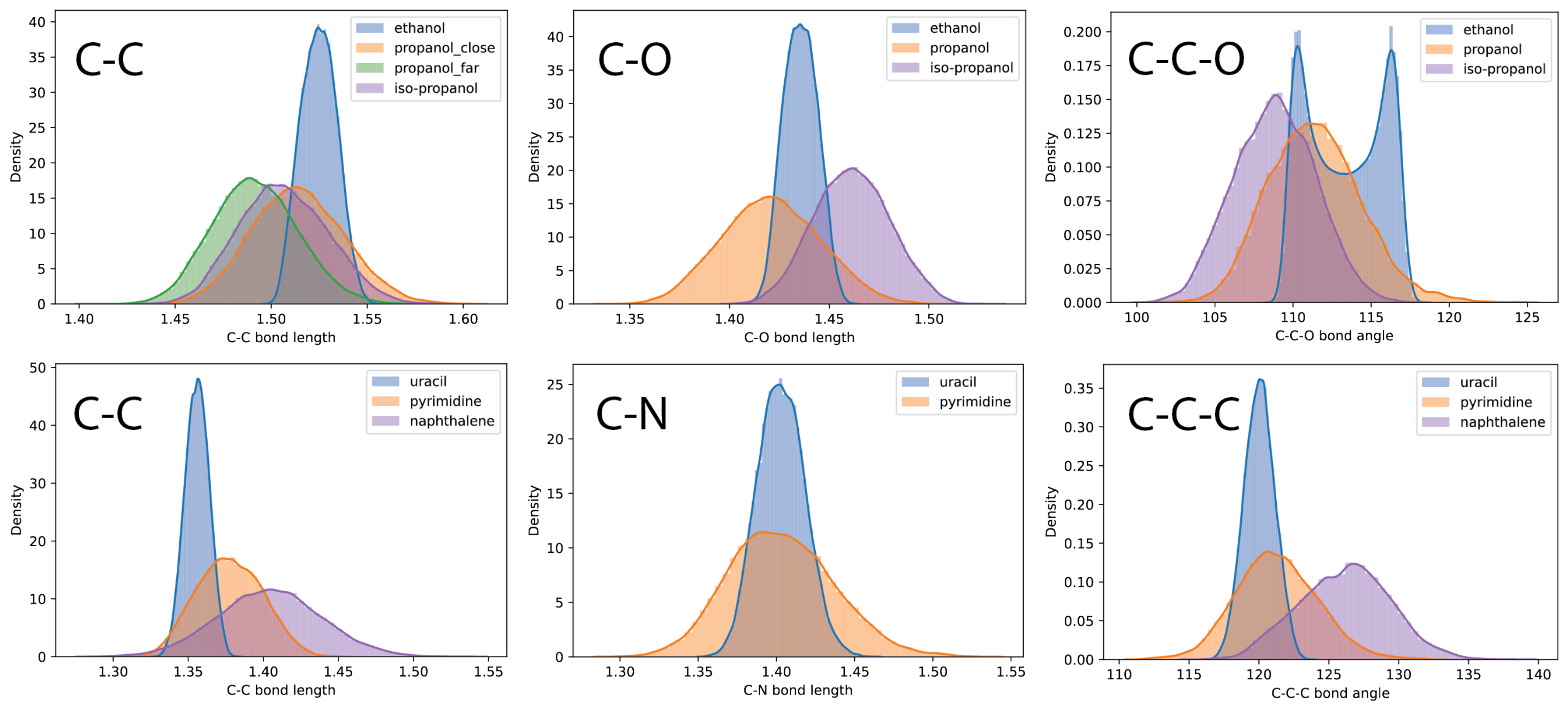}
\caption{Top row, distributions of select bond lengths and angles in ethanol, \textit{n}-propanol, and \textit{iso}-propanol. Bottom row, distributions of select bond lengths and angles in uracil, pyrimidine, and naphthalene. Propanol\_close and propanol\_far of C-C bond correspond to bonds that are close and far away from the oxygen atom in propanol.}
\label{fig:torchmdnet}
\end{figure*}

In Figure ~\ref{fig:molecular_graph}, within each set, the \texttt{TorchMD-NET} model trained with MD trajectories of the molecule on the left was used to perform MD simulations of the molecules on the right. The NVE ensemble was used, where the total number of particles in the simulation box and the box volume are fixed, and the total energy is conserved. For all molecules, 
the timestep and the total simulation time were chosen to be 0.1\,fs 
and 10\,ps (100,000 timesteps), respectively. Figure ~\ref{fig:torchmdnet} shows that the C-C and C-O bond length distributions of ethanol, propanol and \textit{iso}-propanol have similar means, whereas the latter two have a larger spread. It is worth noting that for propanol, the C-C bond length closer to the oxygen atom has a similar distribution to that of ethanol, which is expected because their local environments are similar. For bond angles, The C-C-O bond angle distribution of ethanol exhibits two peaks, whereas the other two only have one peak. For uracil, pyrimidine and naphthalene, the
C-C bond length distribution of naphthalene is shifted to a higher range compared to the other two, which may be due to the absence of N atoms in its ring structure. The C-N bond length distributions of uracil and pyrimidine have similar means, 
whereas the latter has a larger spread. Similarly, we observe comparable C-C-C bond angle distributions in uracil and pyrimidine, whereas the same distribution for naphthalene is shifted to a higher range, again an effect which may be due to the absence of N atoms in naphthalene's ring structure. The similarity and differences of bond length and angle distributions demonstrate
that \texttt{TorchMD-NET} trained on one type of molecule is
transferable to other similar molecules.

\vspace{3mm}

\noindent \textit{Key findings:} We present a 
novel application of \texttt{TorchMD-NET}, in which 
this AI model was fine-tuned to describe a given small molecule by accurately predicting its potential energy and forces and perform NVE MD simulations, and then seamlessly used to describe other molecules with different structures, while still capturing physically realistic bond length and angle distributions.

\subsection{Interpretable Inference}
\label{sec:xai}

Here we explore the use of explainable AI and 
dimension reduction techniques. 
This is motivated by 
recent studies in which explainable AI approaches, such 
as Excitation Backpropagation~\cite{zhang2018top}, CAM~\cite{zhou2016learning} and Grad-CAM~\cite{selvaraju2017grad} and Contrastive gradient~\cite{tieleman2008training}, were used to identify 
key functional groups in molecular structures which 
contribute to toxicity~\cite{pope2018discovering}. 
Similarly, dimension reduction has been used to 
visualize the distribution shift of sampled molecules in the feature space with or without transfer learning\cite{moret2020generative}. We explore 
the use of these tools to gain new insights into the 
information that our AI suite extracts from input data 
to produce reliable predictions.

\begin{figure*}[htbp]
    \centerline{
    \includegraphics[scale=0.4]{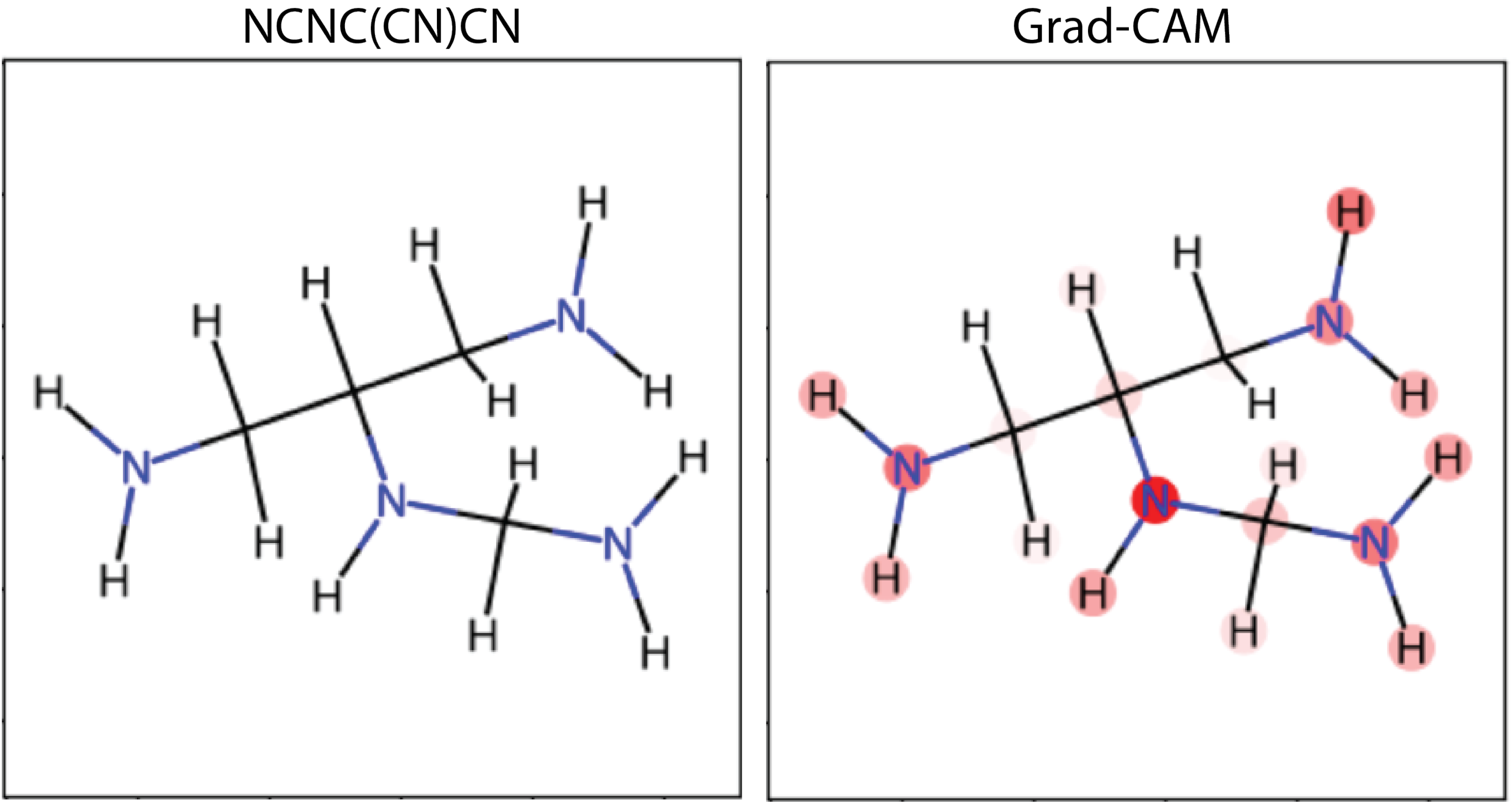}
    }
    \vspace{1.5mm}
    \centerline{
    \includegraphics[scale=0.52]{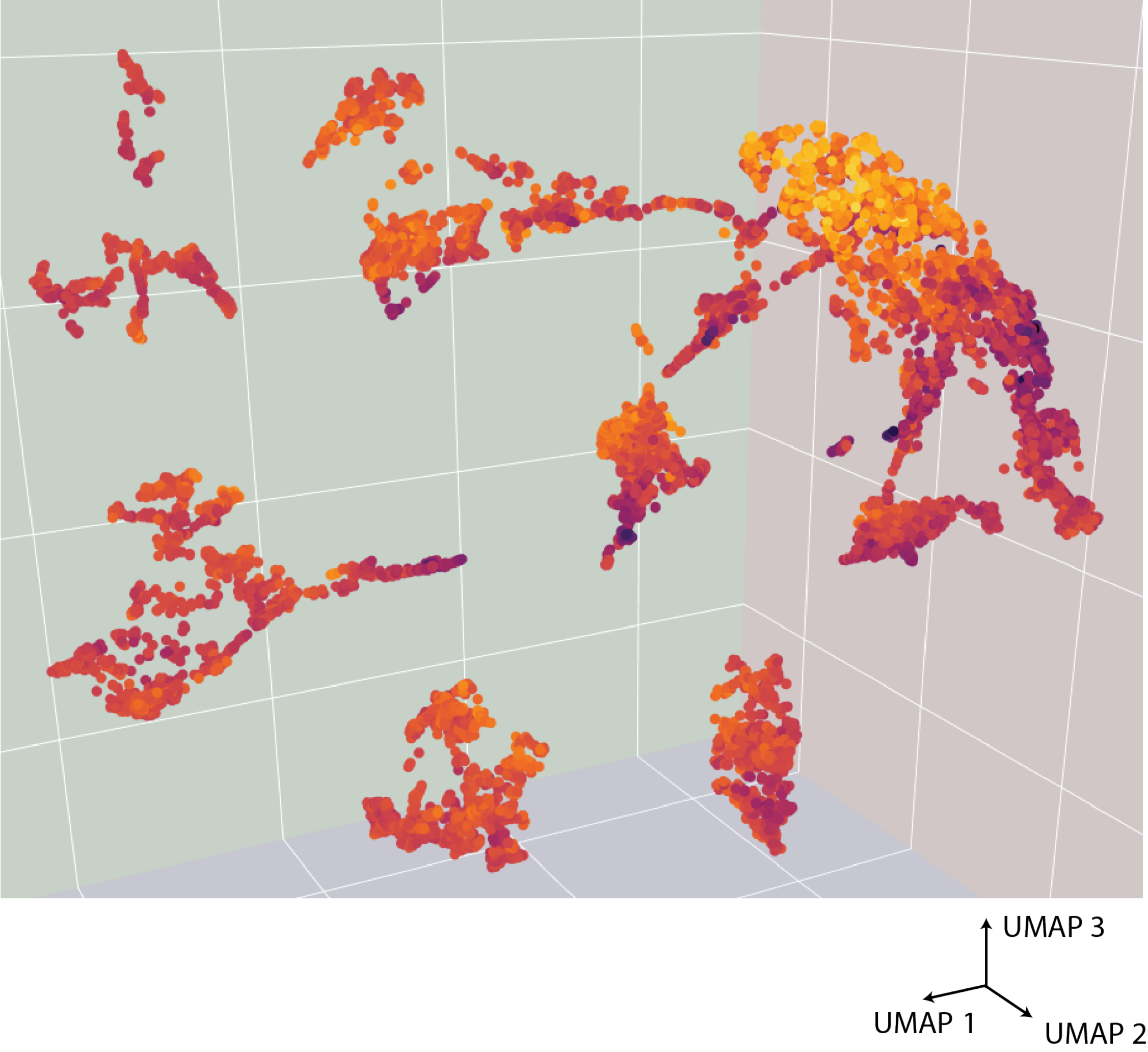}
    }
    \caption{Top panels: exploration of the \texttt{PhysNet} 
    model prediction using Grad-CAM method. The N atoms and the H atoms connected to them are highlighted in red, indicating that they carry more weight in 
    \texttt{HOMO} prediction. Bottom panel: UMAP dimension reduction results for the
    \texttt{PhysNet} model with \texttt{HOMO} as 
    target property. Gold and purple dots represent molecules with 
    high and low \texttt{HOMO} values, respectively.}
    \label{fig:grad-cam}
\end{figure*}

\textbf{Model performance attribution.}
By projecting the second last layer's high dimensional 
vector representation of graph neural network onto 
molecular structure and visualizing the projection, 
we can better understand the
physical and chemical properties of the
input data that affect predictions of our AI models. 
The top panels of Figure ~\ref{fig:grad-cam} present 
model interpretation results of 
how \texttt{PhysNet} model predicts \texttt{HOMO} based on 
molecular structures 
via the Grad-CAM method. The N atoms and the H atoms 
connected to them are highlighted,
possibly indicating that for \texttt{PhysNet}, these atoms
carry more weight 
in the prediction of \texttt{HOMO}. We do not claim that explanations 
found by our deep learning model are the definite reasons for 
accurate prediction of QM properties such as \texttt{HOMO} or 
\texttt{ZPVE}, since these QM properties may not be simply determined by
atomic species and coordinates. However, AI-explained 
visualization can help us better make sense of the patterns of 
complex molecular property predictions. 

\textbf{Dimension Reduction} 
To reveal the correlation between model features and 
target properties, 
we applied the UMAP dimension reduction technique to find the 
distributions of small molecules by projecting their high-dimensional 
structural and chemical data onto low-dimension spaces. 
UMAP has been shown to achieve comparable or even better performance than
other dimension reduction techniques such as principal component analysis (PCA) 
~\cite{jolliffe2016principal} and t-distributed stochastic 
neighbor embedding (tSNE)~\cite{van2008visualizing} on non-linear datasets~\cite{umap}.
Dimension reduction results for \texttt{PhysNet} model 
with \texttt{HOMO} as target 
property is shown in the bottom panel of 
Figure ~\ref{fig:grad-cam}, where each dot in the 
plot represents a molecule, color-coded based on its
\texttt{HOMO} value. A 10\% randomly selected subset (13.4k molecules) of the
\texttt{QM9} dataset was used to produce these results, which consists of stable small
molecules composed of CHONF. 
From the scatter plot we know that molecules with similar \texttt{HOMO} values 
are clustered together and there is clear separation of molecules 
with low and high \texttt{HOMO} values. We present additional 
illustrative results in~\ref{sec:ap_xai}.

\begin{figure}[!htbp]
    \centering
    \includegraphics[scale=0.3]{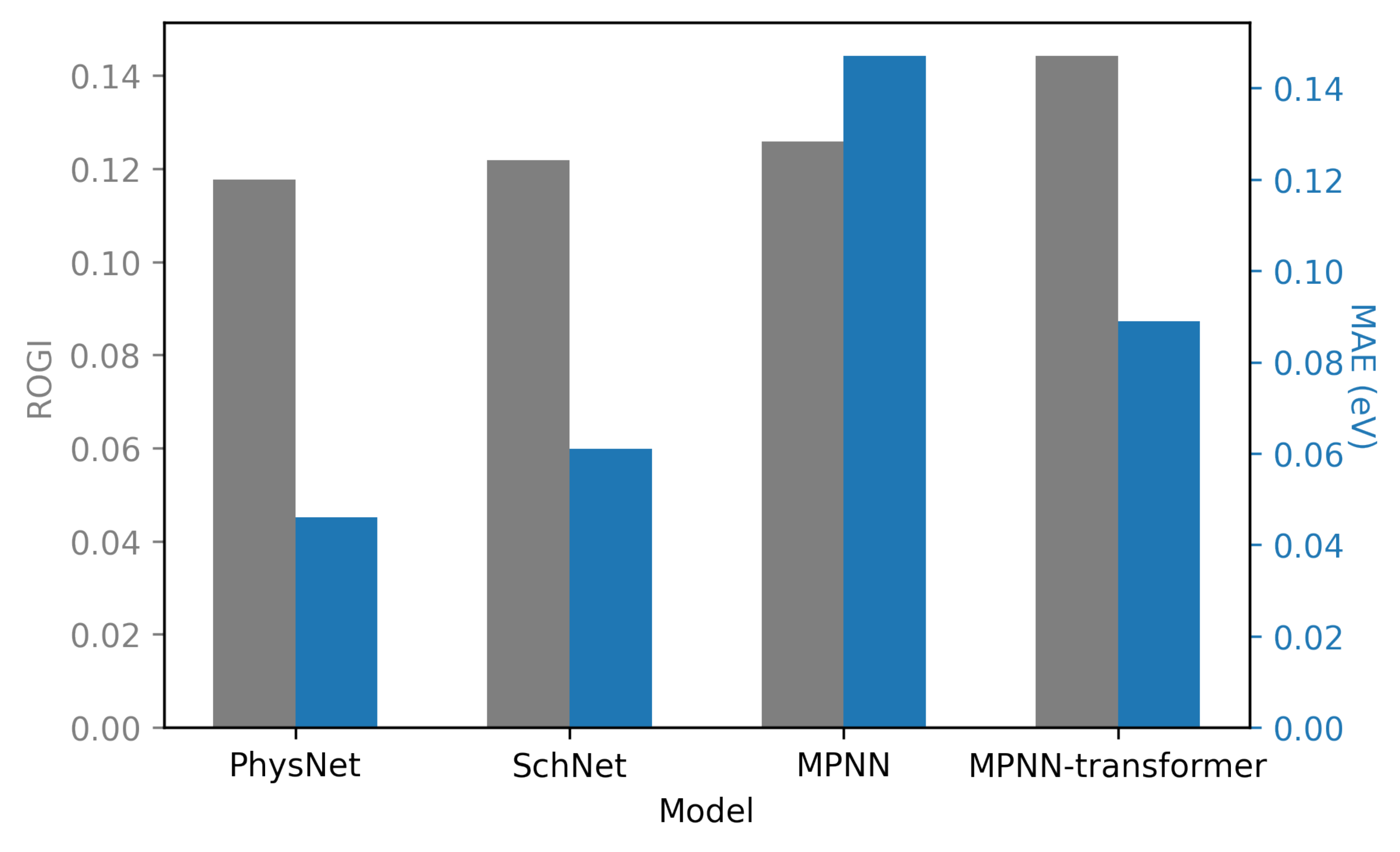}
    \caption{Roughness of the \texttt{HOMO} landscape (gray) and mean absolute error of model predictions (blue) for \texttt{PhysNet}, \texttt{SchNet}, \texttt{MPNN} and \texttt{MPNN-transformer}.}
    \label{fig:rogi}
\end{figure}

\textbf{Roughness of Molecular Property Landscape.}
For molecular property prediction, the predictive performance of 
graph neural networks has been shown to correlate to the roughness 
of molecular property landscape ~\cite{SARI}~\cite{SALI}~\cite{MODI}. We adopted the recently proposed state-of-the-art roughness index (ROGI)~\cite{rogi} to measure
how rough the \texttt{HOMO} and \texttt{ZPVE} landscapes are for four graph neural network models: \texttt{PhysNet}, \texttt{SchNet}, \texttt{MPNN}, \texttt{MPNN-transformer}. The calculation of molecular landscape roughness involves specifying a molecular representation and a distance metric. Molecular representations can be either learned by the graph neural network, with values extracted from the second last layer, or calculated based on molecular structures, as represented by SMILES strings or 3D Cartesian coordinates. The distance metrics are used to measure how different two molecular representations are. Example distance metrics include Tanimoto similarity, Euclidean distance, cityblock distance and cosine distance. To calculate the ROGI values, learned molecular representation and one of the aforementioned distance metrics are used. Using Euclidean distance as the distance metric and \texttt{HOMO} as the target property, we show the ROGI values and mean absolute errors of four graph neural networks in Figure \ref{fig:rogi}.
We observe that a lower ROGI value in general corresponds to a lower mean absolute error, that is, higher predictive performance. We attribute this trend to the direct relation of a higher performing model to the smoothness of the resulting molecular property landscape.
The exception is \texttt{MPNN}, which corresponds to a lower ROGI value despite a higher MAE as compared to the \texttt{MPNN-transformer}, which may be because the addition of transformer layers roughens the molecular property landscape while
facilitating model training.

\vspace{3mm}

\noindent \textit{Key findings:} Our proposed approach 
brings together disparate 
interpretability AI tools to explore and make sense of 
AI model predictions, encompassing model performance 
attribution and scientific visualization; dimension reduction 
with UMAP to explore clustering of molecules with 
similar properties; and metrics such as the roughness index 
to quantify the predictive performance of our AI models 
for QM properties. These complementary tools provide 
valuable insights into the features and patterns of 
input data that are relevant for AI inference.

\section{Conclusion}

The rise of AI in the early 2010s was possible 
by a combination of elements, including 
disruptive technologies and computing approaches, 
as well as the desire to advance state-of-the-art 
practice through collaborative and friendly 
competitions in which high-quality datasets and AI 
models were freely shared. Similar approaches have been 
mirrored in science 
and engineering in recent years. These 
efforts are now being formalized through FAIR 
(findable, accessible, interoperable and reusable) 
initiatives~\cite{fairmetrics,fairguiding} in the context 
of scientific datasets~\cite{fair_hbb_dataset}, 
research software~\cite{fair_rs} and 
AI models~\cite{2022arXiv220700611R,HEP_FAIR_Duarte}. 
This study represents yet another significant step in 
this direction. We have assembled benchmark datasets, 
added novel features to state-of-the-art graph neural 
networks and transformer models, coupled them 
with robust libraries for hyperparameter tuning 
to improve 
their capabilities for scientific discovery, and developed 
and adapted a set of visualization and 
interpretability tools to make sense of the AI predictions. 
All these elements are unified within a single computational 
framework that has been deployed and extensively tested 
on leadership-class, high-performance computing platforms. 
Researchers using this computational framework 
will be able to conduct scientific discovery combining 
state-of-the-art AI models with datasets that are coupled 
with advanced supercomputing platforms. We expect that this approach 
will catalyze the sharing of AI knowledge and tools 
in the context of molecular and crystal property prediction applications.

\section{Acknowledgments}

This work was supported by the FAIR Data program and the 
Braid project of the U.S. Department of Energy, Office of 
Science, Advanced Scientific Computing Research, under 
contract number DE-AC02-06CH11357. It used resources of the 
Argonne Leadership Computing Facility, which is a DOE Office 
of Science User Facility supported under Contract 
DE-AC02-06CH11357. This work was supported by Laboratory 
Directed Research and Development (LDRD) funding from Argonne 
National Laboratory, provided by the Director, Office 
of Science, of the U.S. Department of Energy under 
Contract No. DE-AC02-06CH11357. This research used the 
Delta advanced computing and data resource which is 
supported by the National Science Foundation (award 
OAC 2005572) and the State of Illinois. Delta is a 
joint effort of the University of Illinois at
Urbana-Champaign and its National Center for 
Supercomputing Applications. We thank Prasanna Balaprakash 
and the 
\texttt{DeepHyper} 
team for their expert support and guidance as we 
coupled their library into our computational AI framework. 

\section*{Code availability} 
The AI models presented in this article 
are open source 
in GitLab~\cite{gitlab_code_md}. We also provide a stand alone 
Colab tutorial~\cite{colab_tutorial_md} that shows users how 
to use our framework, i.e., loading data and AI models, 
doing AI inference, and interpreting AI model predictions. 
Following best practices, the datasets used in these 
studies are 
published in Zenodo~\cite{hyun_park_2023_7758490}. 
We provide an exemplary case of our published AI 
suite in~\ref{sec:colab_code}.

\section*{ORCID IDs}
Santanu Chaudhuri \href{https://orcid.org/0000-0002-4328-2947}{0000-0002-4328-2947}\\
Eliu Huerta \href{https://orcid.org/0000-0002-9682-3604}{0000-0002-9682-3604} \\
Hyun Park \href{https://orcid.org/0000-0001-5550-5610}{0000-0001-5550-5610}\\
Emad Tajkhorshid \href{https://orcid.org/0000-0001-8434-1010}{0000-0001-8434-1010}\\
Ruijie Zhu \href{https://orcid.org/0000-0001-9316-7245}{0000-0001-9316-7245}

\bibliographystyle{iopart-num}
\bibliography{sample}

\clearpage
\appendix

\section{Hyperparameter optimization results of PhysNet with HOMO as target property}

\begin{table}[htbp]
    \centering
    \small
    \caption{Top 10 \texttt{DeepHyper} hyperparameter combinations for \texttt{PhysNet} with \texttt{HOMO} as target property.}
    \begin{tabular}{llll}\\
    \hline
\texttt{agb} & \texttt{amp} & \texttt{batch\_size} & \texttt{gradient\_clip}  \\ \hline
3  & TRUE  & 235 & 1.36E-01 \\
1  & TRUE  & 349 & 1.10E+00 \\
14 & FALSE & 130 & 5.40E-05 \\
3  & FALSE & 159 & 1.79E-02 \\
5  & TRUE  & 404 & 8.24E-02 \\
4  & TRUE  & 460 & 7.21E-02 \\
13 & TRUE  & 160 & 3.42E-05 \\
4  & FALSE & 147 & 6.16E-02 \\
7  & TRUE  & 163 & 1.90E-01 \\
1  & TRUE  & 258 & 3.58E-02 \\
\hline
    \label{tab:params_ap1}
    \end{tabular}
\end{table}

\begin{table}[htbp]
    \centering
    \small
    \caption{As Table~\ref{tab:params_ap1} for the rest of parameters optimized through \texttt{DeepHyper}.}
    \begin{tabular}{llll}\\
    \hline
\texttt{learning\_rate} & \texttt{optimizer}   & \texttt{weight\_decay} & \texttt{objective}  \\
\hline
1.15E-03 & sgd          & 2.94E-05 & -1.604 \\
8.69E-04 & sgd          & 2.30E-06 & -2.454 \\
1.49E-04 & lamb         & 1.17E-03 & -2.976 \\
5.64E-01 & lamb         & 2.09E-04 & -3.323 \\
1.38E-03 & sgd          & 4.83E-05 & -6.474 \\
2.33E-02 & sgd          & 2.84E-04 & -7.214 \\
4.61E-04 & lamb         & 1.74E-04 & -12.494 \\
5.99E-04 & torch adamw & 1.59E-03 & -14.5111 \\
5.26E-01 & lamb         & 1.44E-04 & -18.0449 \\
6.95E-01 & lamb         & 8.62E-05 & -29.505 \\
\hline
    \label{tab:params_ap2}
    \end{tabular}
\end{table}

\section{Examples of model explainability features}
\label{sec:ap_xai}
\begin{figure}[htbp]
    \centering
    \includegraphics[scale=0.52]{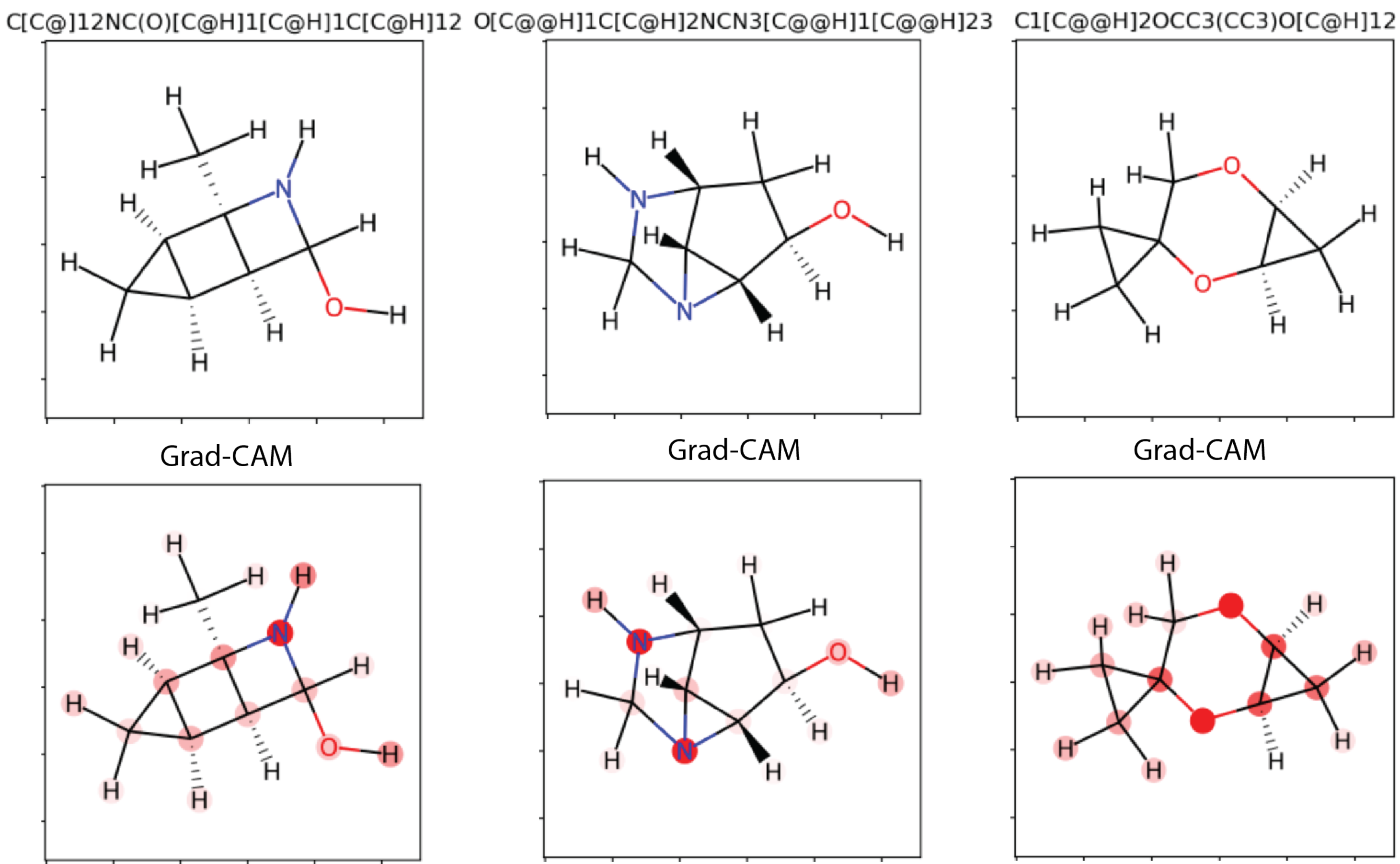}
    \caption{Molecular graphs (top row) and Grad-CAM interpretations (bottom row) of \texttt{PhysNet} with \texttt{HOMO} as target property for three example molecules. The N atoms and the H atoms attached to them are highlighted in red, indicating that they carry more weight in model predictions.}
    \label{fig:apx_b1}
\end{figure}

We present results to complement the interpretable AI 
analysis presented in Section~\ref{sec:xai}. 
Figure ~\ref{fig:apx_b1} illustrates what information AI 
models may extract from input data to make predictions 
that are consistent with state-of-the-art knowledge on 
QM properties.  

\subsection{Grad-CAM interpretation}

\pagebreak

\subsection{UMAP interpretation}
\begin{figure}[htbp]
    \centerline{
    \includegraphics[scale=0.34]{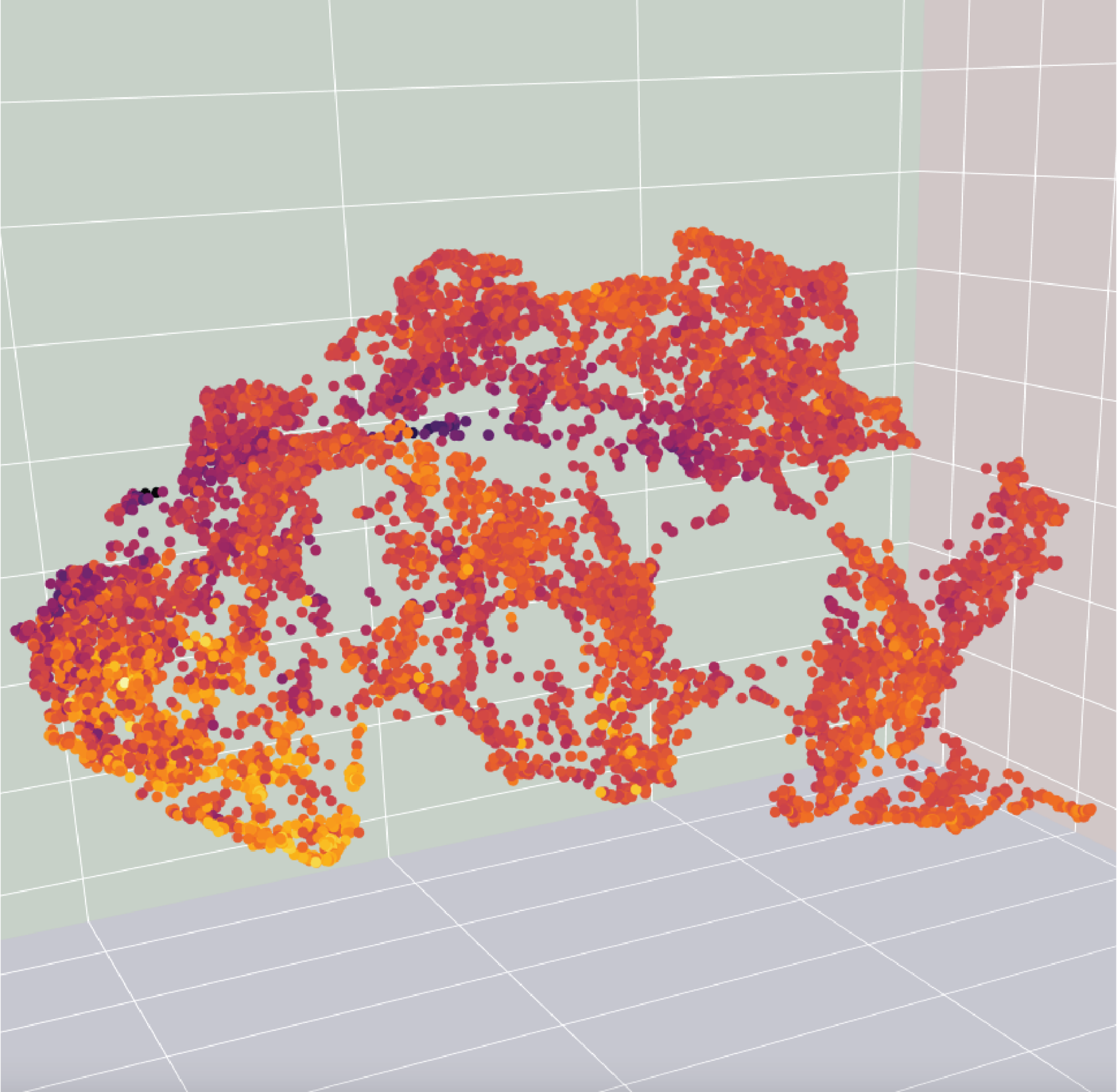}
    \includegraphics[scale=0.34]{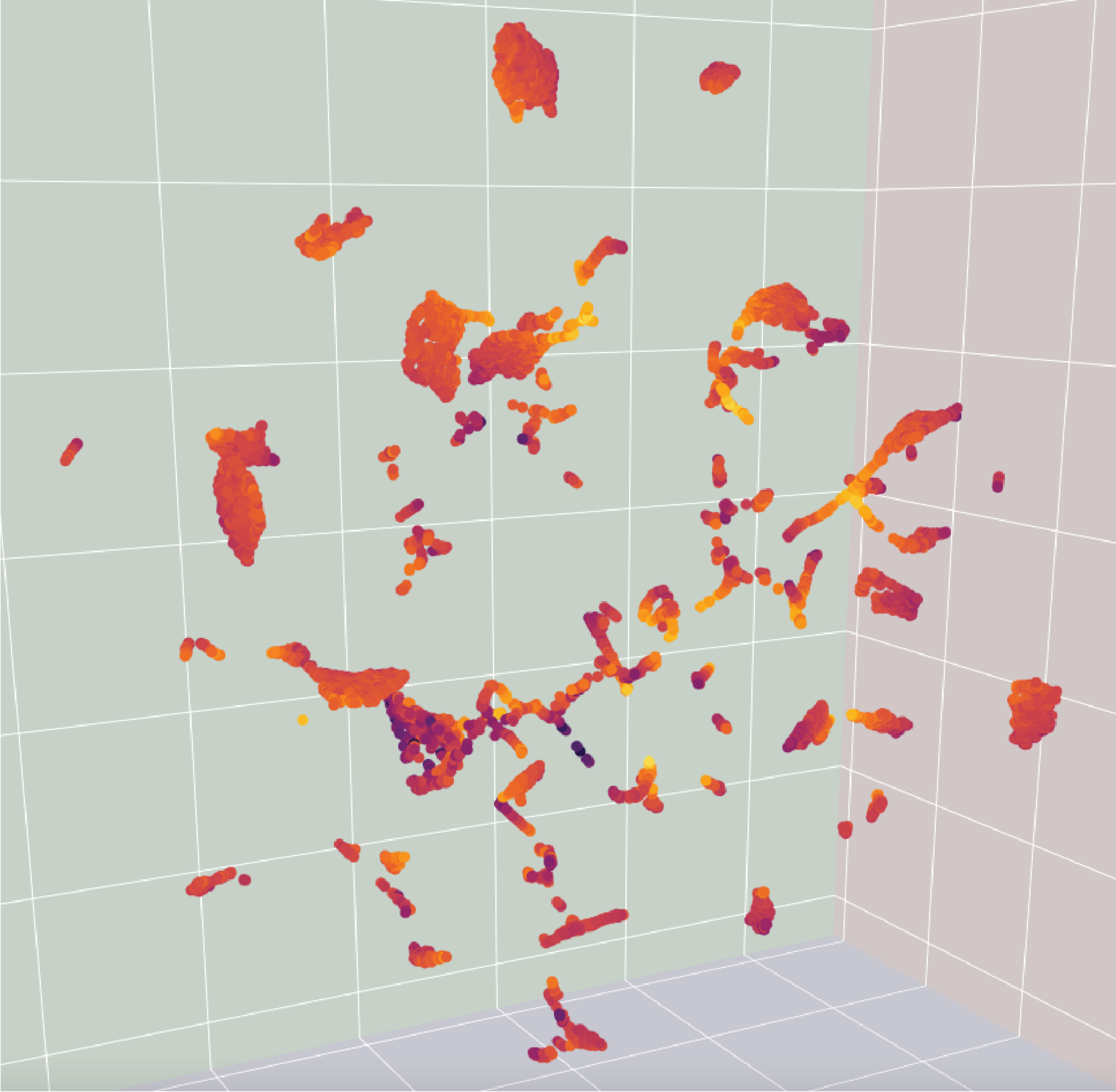}
    }
    \vspace{1.4mm}
    \centerline{
    \includegraphics[scale=0.34]{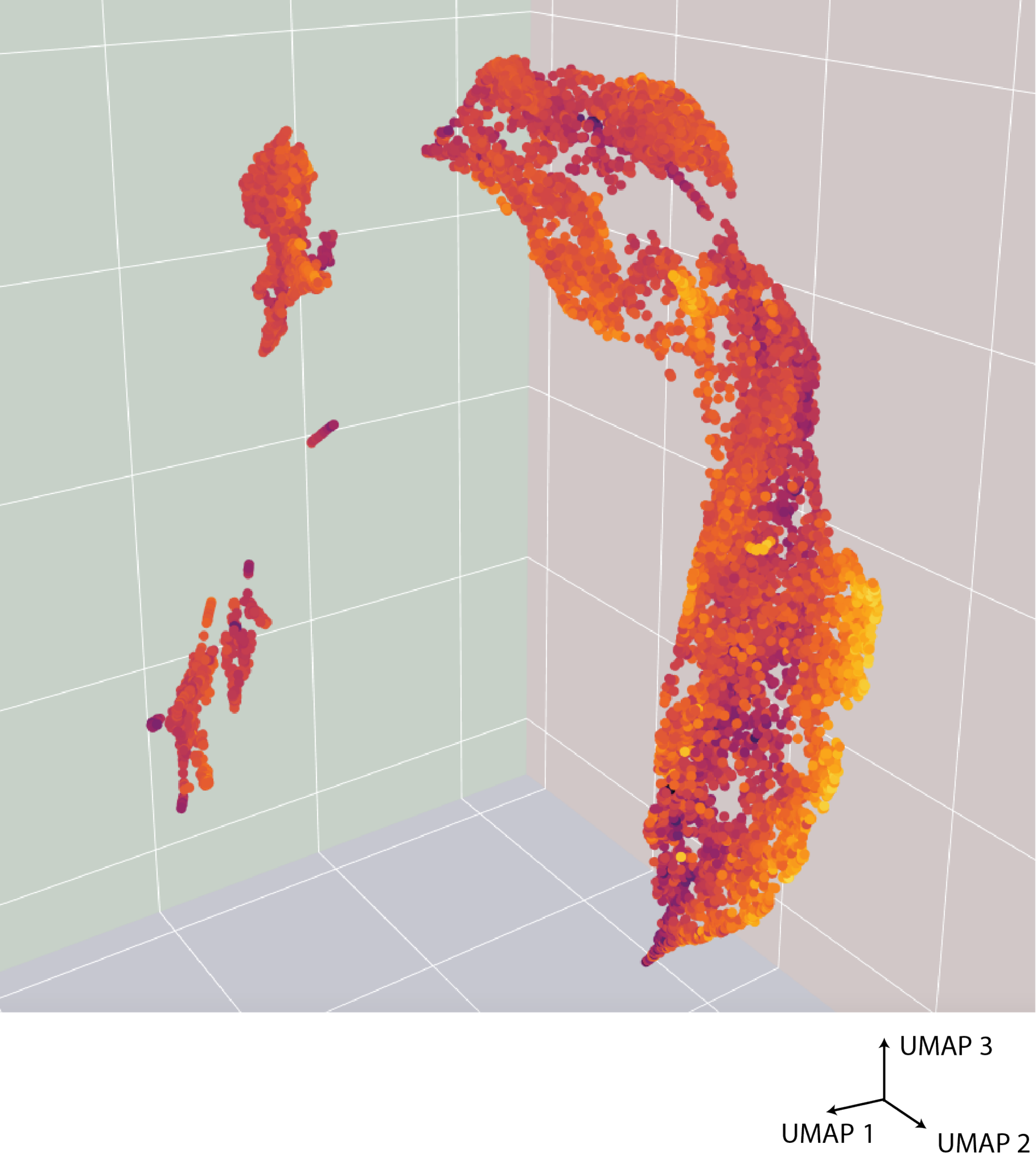}
    }
    \caption{UMAP dimension reduction results for \texttt{SchNet} (top left panel),  \texttt{MPNN} (top right panel) 
    and \texttt{MPNN-transformer} (bottom panel) with \texttt{HOMO} as target property. A randomly selected 10\% subset (13.4k molecules) of the \texttt{QM9} dataset was used for analysis, which consists of stable small organic molecules composed of CHONF.}
    \label{fig:apx_b2}
\end{figure}

\noindent Figure ~\ref{fig:apx_b2} shows that we can turn 
our AI predictors into feature extractors to explore clustering of molecules
with similar properties. 

\pagebreak

\subsection{Example code}
\label{sec:colab_code}
Here we provide an exemplar of our 
released AI suite~\cite{colab_tutorial_md}. The 
scientific software below shows, step-by-step, how 
to download the \texttt{hMOF} dataset that we published 
in Zenodo~\cite{hyun_park_2023_7758490}, and 
then how to use our \texttt{CGCNN} model to infer $\textrm{CO}_2$ adsorption for a variety of 
MOF structures.

\begin{lstlisting}[language=Python, caption={Exemplar of our step-by-step tutorials, published 
in~\cite{colab_tutorial_md}, which shows how to combine the \texttt{hMOF} dataset and our 
\texttt{CGCNN} model to compute \(\textrm{CO}_2\) adsorption for a variety of MOF structures.}]

#Below pip install all necessary packages
!pip install torch torchvision torchaudio --extra-index-url https://download.pytorch.org/whl/cu116
!pip install pyg_lib torch_scatter torch_sparse torch_cluster torch_spline_conv torch_geometric -f https://data.pyg.org/whl/torch-1.13.0+cu116.html
!pip install pytorch-lightning
!pip install  dgl -f https://data.dgl.ai/wheels/cu116/repo.html
!pip install  dglgo -f https://data.dgl.ai/wheels-test/repo.html
!pip install rdkit
!pip install einops curtsies p_tqdm transformers pathlib scikit-image argparse wandb cairosvg h5py pynvml jupyter
!pip install persim ripser ray PyCifRW fairscale ase fast-ml
!pip install pymatgen timm captum
!pip install MDAnalysis 
!pip install scikit-tda umap-learn plotly dash

#Below import relevant packages
import torch, torch_geometric, torch_cluster, pytorch_lightning, dgl, ray, rdkit, ripser, sklearn, transformers, timm, h5py, cairosvg, ase, os, sys, pymatgen, os

#Below set variable names
ROOT="/content"
os.chdir(ROOT)
DATA_ROOT=os.path.join(ROOT, "data") 
CRYSTAL_DATA_ROOT=os.path.join(ROOT, "hMOF/cifs") 
SAVE_ROOT=os.path.join(ROOT, "pretrained_models") 
os.makedirs(SAVE_ROOT, exist_ok=True)
os.makedirs(DATA_ROOT, exist_ok=True)

#Below get tar-zipped gitlab repository
!wget https://zenodo.org/record/7758490/files/ai4molcryst_argonne-0.0.1.tar.gz?download=1 -O ai4molcryst_argonne-0.0.1.tar.gz
!tar -xvf ai4molcryst_argonne-0.0.1.tar.gz
!ls ai4molcryst_argonne-0.0.1/

#Below move some necessary files
os.chdir("ai4molcryst_argonne-0.0.1")
!mv main/main_pub.py .
!mv main/config_pub.py .
!mv main/dlhub.py .
os.system("mv dlhub.py dlhub.yaml")

#Below import modules and define dictionary
import main_pub
import config_pub
from train.dist_utils import *
from main_pub import *
import yaml

with open("dlhub.yaml","r") as f:
    d = yaml.safe_load(f)
class obj(object):
    """dict to object"""
    def __init__(self, d):
        for k, v in d.items():
            if isinstance(k, (list, tuple)):
                setattr(self, k, [obj(x) if isinstance(x, dict) else x for x in v])
            else:
                setattr(self, k, obj(v) if isinstance(v, dict) else v)
opt = obj(d)

#Below download hMOF dataset
os.chdir(ROOT)
!wget https://zenodo.org/record/7758490/files/cifs.zip?download=1 -O cifs.zip
!unzip cifs.zip
os.chdir("ai4molcryst_argonne-0.0.1")

#Below download pretrained CGCNN model from Zenono
!mkdir pretrained_models
!wget https://zenodo.org/record/7758490/files/cgcnn_pub_hmof_5000.pth?download=1 -O $SAVE_ROOT/cgcnn_pub_hmof_5000.pth
ckpt = torch.load(os.path.join(SAVE_ROOT, 'cgcnn_pub_hmof_5000.pth'), map_location=torch.device('cpu'))
model = config_pub.BACKBONES["cgcnn"](**config_pub.BACKBONE_KWARGS["cgcnn"])
print(model.__class__.__name__)
model.load_state_dict(ckpt['model'])

#Below define dataloader
opt.log = False
opt.backbone = "cgcnn"
opt.gpu = True
opt.name = "cgcnn_pub_hmof_5000"
opt.epoches = 1000
opt.batch_size = 16
opt.optimizer = "torch_adam"
opt.data_dir = DATA_ROOT
opt.data_dir_crystal = CRYSTAL_DATA_ROOT
opt.use_tensors = True
opt.load_ckpt_path = SAVE_ROOT
opt.dataset = "cifdata"
opt.task = "homo"
opt.which_mode = "infer"
opt.smiles_list = None
opt.crystal = True
opt.num_oversample = 0
train_loader, val_loader, test_loader, mean, std = call_loader(opt) #Loader and data sets

#Below define inference
import tqdm
def one_inference(dataloader: torch.utils.data.DataLoader):
    device="cpu" #torch.cuda.current_device()
    model.to(device)
    model.eval()
    return_preds = collections.defaultdict(list)
    props, diffs, targs = [], [], []

    for pack, _ in tqdm.tqdm(dataloader, total=len(dataloader)):
        atom_fea, nbr_fea, nbr_fea_idx, crystal_atom_idx, batch, dists, targetE = pack.x, pack.edge_attr, pack.edge_index, pack.cif_id, pack.batch, pack.edge_weight, pack.y
        pack =  atom_fea, nbr_fea, nbr_fea_idx, batch, dists, targetE

        atom_fea, nbr_fea, nbr_fea_idx, batch, dists, targetE = to_cuda(pack) if device != "cpu" else pack
        pred=model.forward(atom_fea.to(device), nbr_fea.to(device), nbr_fea_idx.to(device), dists.to(device), crystal_atom_idx.to(device), batch.to(device))
        props.append(pred.view(-1).detach().cpu().numpy())        targs.append(targetE.view(-1).detach().cpu().numpy())        diffs.append(pred.view(-1).detach().cpu().numpy() - targetE.view(-1).detach().cpu().numpy())

    return_preds["property"] = np.concatenate(props, axis=0)
    return_preds["targets"] = np.concatenate(targs, axis=0)
    return_preds["property_difference"] = np.concatenate(diffs, axis=0)
    return return_preds

#Below run inference
def run_inference(input_data):
    predictions: dict = one_inference(input_data)
    return predictions
preds = run_inference(test_loader)

\end{lstlisting}

\pagebreak

\end{document}